\documentclass[prx,aps,twocolumn,showpacs]{revtex4-1}

\usepackage{amssymb,amsmath}
\usepackage{graphicx}
\usepackage{multirow}
\usepackage{color}
\usepackage[english]{babel}
\usepackage{hyperref}
\hypersetup{
    colorlinks=true,
    linkcolor=blue,
    filecolor=blue,      
    urlcolor=blue,
    citecolor=blue
}
\usepackage{array}

\newcommand{\be}{\begin{equation}}
\newcommand{\ee}{\end{equation}}
\newcommand{\ba}{\begin{equation}}
\newcommand{\ea}{\end{equation}}
\newcommand{\bea}{\begin{eqnarray}}
\newcommand{\eea}{\end{eqnarray}}

\newcommand{\w}{\omega}

\newcommand{\tu}{\tilde{U}}
\newcommand{\bq}{\textbf{q}}
\newcommand{\br}{\textbf{r}}
\newcommand{\te}{\tilde{\epsilon}}

\newcommand{\tz}{\tilde{z}} 
\newcommand{\bg}{\bold{G}}

\newcommand{\e}{\epsilon}
\newcommand{\Ai}{\mathrm{Ai}}

\renewcommand{\Re}{\mathrm{Re}}

\newcommand{\eref}[1]{Eq.~(\ref{#1})}
\newcommand{\esref}[1]{Eqs.~(\ref{#1})}

\newcommand{\ocite}[1]{Ref.~\cite{#1}}

\begin{document}
\title{Switching between relaxation hotspots and  coldspots in disordered spin qubits}
\author{Amin Hosseinkhani}
\email{amin.hosseinkhani@uni-konstanz.de}
\affiliation{Department of Physics, University of Konstanz, D-78457 Konstanz, Germany}
\author{Guido Burkard}
\email{guido.burkard@uni-konstanz.de}
\affiliation{Department of Physics, University of Konstanz, D-78457 Konstanz, Germany}

\begin{abstract}
We develop a valley-dependent envelope function theory that 
can describe the effects of
arbitrary configurations of interface steps and miscuts on the qubit relaxation time. For a given interface roughness, we show how our theory can be used to find the valley-dependent dipole matrix elements, the valley splitting, and the spin-valley coupling as a function of the electromagnetic fields in a Si/SiGe quantum dot spin qubit. We demonstrate that our theory can quantitatively reproduce and explain the result of experimental measurements for the spin relaxation time with only a minimal set of free parameters. 
Investigating the sample dependence of spin relaxation, we find that at certain conditions for a disordered quantum dot, the spin-valley coupling vanishes. This, in turn, completely blocks the valley-induced qubit decay. We show that the presence of interface steps can in general give rise to a strongly anisotropic behavior of the spin relaxation time. 
Remarkably, by properly tuning the gate-induced out-of-plane electric field, it is possible to turn the spin-valley hotspot into a ``coldspot" at which the relaxation time is significantly prolonged and where the spin relaxation time is additionally first-order insensitive to the fluctuations of the magnetic field. This electrical tunability enables on-demand fast qubit reset and initialization that is critical for many quantum algorithms and error correction schemes.
We therefore argue that the valley degree of freedom can be used as an advantage for Si spin qubits.
\end{abstract}
	
\maketitle
\section{Introduction}
Silicon quantum dots offer an attractive platform for scalable quantum computing \cite{Zwanenburg13}. Two plausible properties of silicon that make it a suitable host material are the weak spin-orbit interaction as well as the abundance of nuclear zero-spin isotopes. These have enabled achieving long relaxation  \cite{Morello10, Yang2013} and dephasing times \cite{Assali11, Tyryshkin12, Steger12} in individual spin qubits. It has recently been shown that the quantum coherence in silicon spin qubits can be maintained even for high temperatures above one Kelvin \cite{Yang20,Petit20}. In order to scale up and build a quantum network of spin qubits, one promising approach to couple the long-distance spins is via coherent interaction with microwave photons \cite{GB20, MB20}. While the strong coherent spin-photon coupling using superconducting resonators has already been realized \cite{Mi18-1, Samkharadze18}, another possibility to couple spin qubits is to coherently transport them by tuning the electric gates. This so-called spin shuttling is investigated for Si spin qubits from both experimental \cite{Mills19,Yoneda20}  and  theoretical \cite{Ginzel20} perspectives.

While silicon quantum dots enjoy the properties mentioned above, they also suffer from one well-known problematic feature, namely the 6-fold degenerate valley states in bulk silicon. In Si heterostructures and quantum dots, a combination of biaxial strain together with the sharp interface potential lifts the valley degeneracy and gives rise to two low-lying states \cite{Zwanenburg13}.  These two valley states can in principle be used to encode the quantum information \cite{Rohling12,Rohling14,Mi18,Penthorn19,Penthorn20}. However, for spin qubits, the presence of the valley states significantly limits the qubit lifetime when the valley energy splitting becomes comparable to the qubit Zeeman splitting. This spin-valley relaxation hotspot was first experimentally observed in \ocite{Yang13}, and since then, it has been the subject of numerous studies \cite{Huang14, Petit18, Borjans19, Hollmann20, Zhang20, Huang20}.

In order to properly understand the behavior of the spin relaxation induced by the valley states in a Si quantum dot, one needs to have a number of important quantities at one's disposal. These include the valley splitting, the inter-valley and intra-valley dipole matrix elements, and the spin-valley coupling caused by the spin-orbit interaction. While the valley splitting has been thoroughly studied by a number of papers \cite{Friesen07,Koiller09,Friesen10,Koiller11,Culcer12,Boross16,Tariq,Hosseinkhani20}, to our knowledge, the works concerned with analyzing the behavior of the spin relaxation have always postulated that the dipole matrix elements and the spin-valley coupling are finite quantities, and these are then simply treated as fitting parameters without considering their microscopic origin. As we will show in this paper, the valley splitting, the dipole matrix elements and the spin-valley coupling do not explicitly depend on each other. However, they are all strongly influenced by an important common factor: the Si/barrier interface roughness. 

Given the experimental process of fabricating silicon heterostructures, the formation of steps and miscuts at the Si/barrier interface is very probable \cite{Hollmann20, Zandvliet93}. It has been shown that the presence of interface steps can severely suppress the valley splitting \cite{Friesen10,Boross16,Tariq,Hosseinkhani20}. Furthermore, the interface steps generally break the in-plane mirror symmetry and therefore one expects that the in-plane dipole moments in a disordered quantum dot become finite (i.e. nonzero) quantities. To our knowledge, so far, all of the published works that use the effective mass theory in analyzing the valley splitting neglect the corrections to the envelope function due to the valley coupling. We argue in this work that using such a valley-independent envelope function is incapable of determining the inter-valley matrix elements, even in the presence of interface steps.  Here  we develop a \textit{valley-dependent} envelope function theory in the presence interface disorder. This theory then enables us to calculate and analyze the mentioned important quantities; namely, the valley splitting, the inter-valley and intra-valley dipole matrix elements, and the spin-valley coupling. We can calculate these quantities for an arbitrary configuration for the interface roughness as a function of the lateral size of the quantum dot and the electromagnetic fields. 

While the spin-orbit interaction is relatively weak in bulk silicon, the structural inversion asymmetry and the symmetry breaking due to the Si/barrier interface  lead to both Rashba-like and Dresselhaus-like spin-orbit interaction. Indeed, it has been shown that the spin-orbit interaction occurs locally at the position of the interface, and it quickly vanishes away from the interface  \cite{Golub04,Nestoklon06,Prada11}. As such, the presence of the interface steps, in turn, modifies the interface-induced spin-orbit interaction. In particular, each time a single-layer atomic step is encountered at the interface, the coefficient of the Dresselhaus term must change sign due to the crystal symmetry of silicon \cite{Ferdous18,FerdousPRB18,Jock18}. Here we employ a description for the spin-orbit interaction which is localized at the disordered interface. Our model describes a 3D electron and  takes into account the sign change of the Dresselhaus term that occurs due to the presence of single-layer interface steps.

\begin{figure}[b!]
\begin{center}
\includegraphics[width=0.5\textwidth]{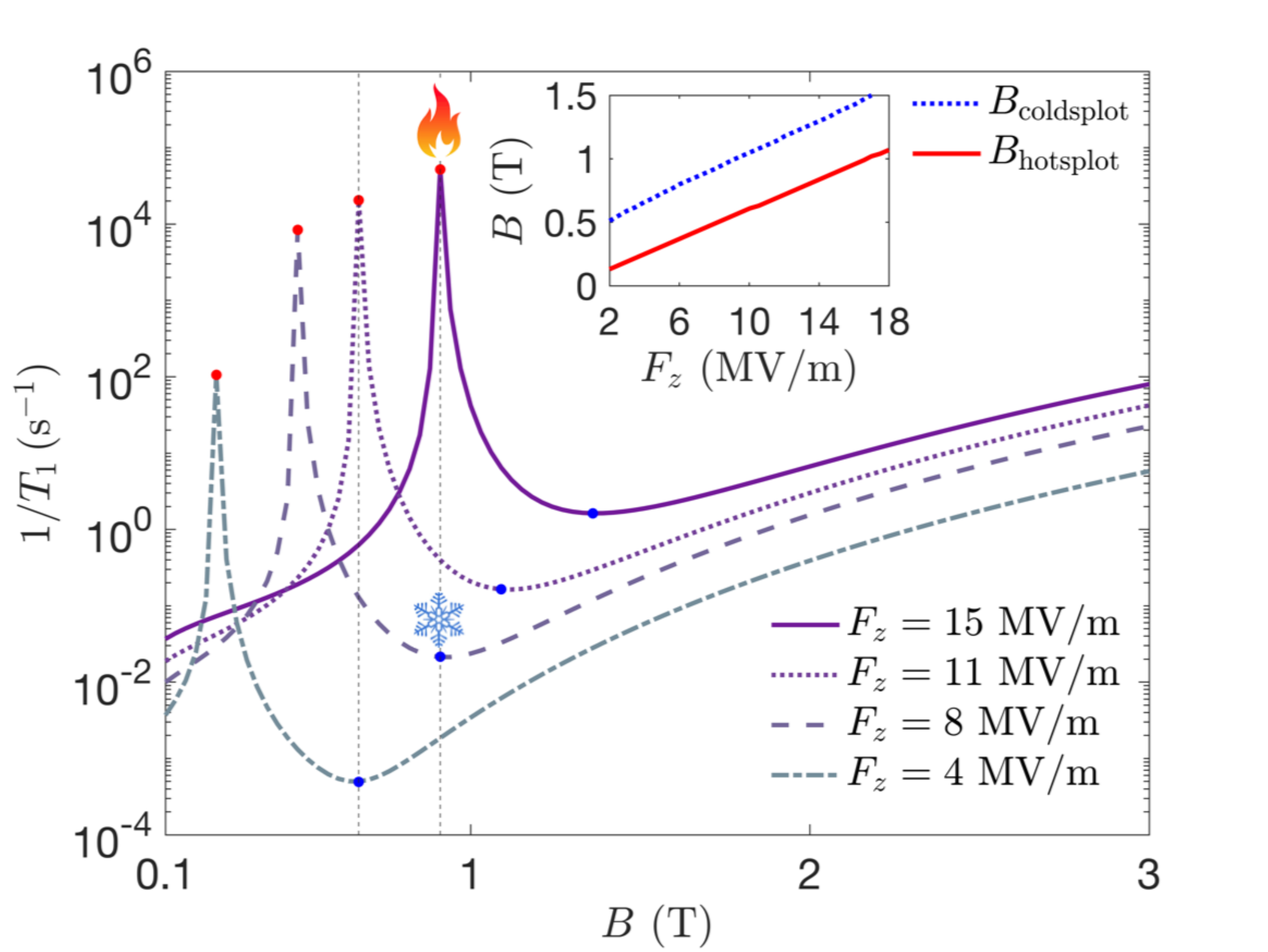}
\end{center}
\caption{The qubit relaxation rate $1/T_1$ as a function of the magnetic field for some fixed values for the out-of-plane electric field. Here for each value of $F_z$, the hotspot (coldspot) is marked by a read (blue) point and it is additionally highlighted by a flame (snowflake) symbol for $F_z=15$ (8) MV/m.  Inset: Magnetic field at which the hotspot and coldspot occurs as a function of the electric field. All the other parameters are the same as given in the caption of Fig.~\ref{fig:T_1_the_vs_exp}.  }  \label{fig:T_1_vs_Fz}
\end{figure}

We  proceed by using our theory to find the modifications to the spin qubit levels due to spin-valley mixing (SVM) and spin-orbit mixing (SOM). This enables us to formulate the theory of spin relaxation in the presence of interface disorder that involves only a minimal number of free parameters. This theory can well describe the spin relaxation for all ranges of the magnetic field including below, around, and above the spin-valley hotspot (Fig.~\ref{fig:T_1_vs_Fz}). We consider a simple model where there are (up to) two interface steps, one to the right  of the quantum dot center and the other to the left of the dot center (Fig.~\ref{fig:dis_dot}).  Using this simple model, we show that our theory for the spin relaxation can quantitatively reproduce the results of experimental measurements presented in \ocite{Borjans19}.
We find that whereas models with a single step cannot quantitatively explain the experimental data around the hotspot, assuming more than two interface steps would give rise to a similar behaviour to what we find with up to two steps.

Having verified that our model can explain the relevant experimental findings, we consider a crossover from a highly disordered quantum dot (where at least one step is very close to the quantum dot center) to a nearly ideal quantum dot (where both steps are away from the quantum dot center) and study the behavior of qubit relaxation time. 
It has been known that the change in the quantum well thickness due to interface steps would give rise to a phase shift that reduces the valley splitting. This renders the spin relaxation time a device-dependent quantity. Here we show that for certain interface roughnesses and electromagnetic fields, the spin-valley coupling vanishes. This has a profound effect on the qubit relaxation time as it completely removes the valley-induced qubit decay, and therefore, in this case the spin-valley hotspot is absent.

As a next step, we investigate how the qubit relaxation time behaves as a function of the out-of-plane electric field, see Fig.~\ref{fig:T_1_vs_Fz}. In our case, this electric field $F_z$ is in turn generated, and can be tuned, by the gate electrodes  surrounding the quantum dot, and it sets the amplitude of the electron wave function at the interface. Therefore the electric field controls the valley splitting, the dipole matrix elements, and the strength of the spin-orbit interaction. As such, changing $F_z$ completely alters the spin qubit levels and, consequently, the qubit relaxation time. Remarkably, we can show that by properly tuning the electric field, the spin-valley hotspot (highlighted in Fig.~\ref{fig:T_1_vs_Fz} by a flame symbol) can dramatically be turned into a ``coldspot" (highlighted by a snowflake symbol) at which the relaxation time is enhanced by several orders of magnitude, and is additionally first-order insensitive to magnetic field fluctuations. While controlling the interface roughness during the fabrication can be a difficult task, tuning the electric field appears to be more achievable. We show that even a small proper change of the electric field can substantially improve the qubit performance by limiting the valley-induced decay. 

On the other hand, electrically tuning the qubit to the spin-valley hotspot is one possible way for a quick initialization to the ground state. The ability to initialize the qubit is one basic criterion for the physical realization of quantum computation \cite{DiVincenzo20}, and it also has a crucial importance in performing quantum error correction \cite{Terhal15}. Indeed, while the valley degree of freedom has so far been viewed as a disadvantage for silicon, we argue that provided the ability to electrically tune the electron wave function at the interface, the valley coupling can serve as a very useful resource for qubit initialization, paving the way towards salable silicon-based quantum information processing.

We finally consider the qubit relaxation as a function of the direction of the magnetic field. Controlling the strength of the spin-orbit interaction via the direction of the B-field is an interesting subject and has been experimentally studied in \ocite{Tanttu19}. Our analysis enables us to thoroughly investigate the anisotropic behavior of the spin-valley coupling. We find that for a disordered quantum dot, the spin relaxation time can have a  strong dependency on the magnetic field direction. This phenomenon was recently noted in an experiment \cite{Zhang20}. Here we provide an alternative explanation by taking into account the effects that emerge when  interface steps are present. 

The remainder of this paper is structured as follows. In Sec.~\ref{sec:VDEnv} we develop the valley-dependent envelope function theory and show how we can calculate the valley splitting and dipole matrix elements. In Sec~\ref{sec:SOI} we discuss the form of symmetrized interface-induced spin-orbit interaction and study how in our model the interface roughness influences the spin-orbit interaction. In Sec.~\ref{sec:SQL} we analyze the modifications to the spin qubit levels due to spin-valley mixing and spin-orbit mixing. We further calculate the spin-valley coupling and study its anisotropic properties. In Sec.~\ref{sec:qr} we present the relations for qubit relaxation rates due to SVM and SOM caused by electron-phonon interaction and the Johnson and 1$/f$ charge noise. In Sec.~\ref{sec:dis} we present our results on the qubit relaxation time and discuss several cases where we investigate the relaxation as a function of interface roughness, direction of the magnetic field, and the electric field. Finally, in Sec.~\ref{sec:conc} we present our conclusions and outlook. The Appendices contain
further details of our analysis.

\section{Valley-dependent envelope function theory}\label{sec:VDEnv}

\subsection{Valley dependent wavefunctions}\label{subsec:wave}
In this section we employ the formalism introduced in \ocite{Friesen07} and build on some of the results and methodology developed in \ocite{Hosseinkhani20} in order to obtain a valley-dependent envelope function in the presence of magnetic field and interface steps. We note that the effects of an in-plane magnetic field as well as interface steps are considered in \ocite{Hosseinkhani20} within the framework of a valley-independent envelope function theory in order to study the valley splitting. However, in the following we show that using a valley-independent envelope function is not sufficient to study the  inter-valley dipole matrix element, as one requires  knowledge of the correction to the envelope function that originates from the coupling between the two valleys.

In the effective mass approximation, the wave function can be written as,
\begin{align}
\label{eq:vowf}
|\nu\rangle=\sum_{j=\pm z} a_je^{ik_jz}u_{j}(r)\Psi_{xyz}^{j},
\end{align}
where $k_{\pm z}=\pm k_0$,  $k_0=0.85\frac{2\pi}{a_0}$, and $a_0=0.543$ nm is the length of the silicon cubic unit cell. Here $u_{\pm z}(r)$ are the periodic parts of the Bloch functions for the $\pm z$ valleys. We can express these functions by a plane wave expansion,
\begin{align}\label{eq:U_r_eq}
u_{\pm z}(r)=\sum_\textbf{G} C_{\pm}(\textbf{G})e^{i\textbf{G}\cdot\textbf{r}},
\end{align}
where the sum runs over
reciprocal lattice vectors $\textbf{G}=(G_x,G_y,G_z)$. The coefficients in this expansion for the two valleys are related via the time-reversal symmetry relation $C_{-}(\textbf{G})=C_{+}^*(-\textbf{G})$. The wave vectors and their corresponding coefficients $C_{+}(\textbf{G})$ for Si are studied and given in \ocite{Koiller11}. $\Psi_{xyz}^{j}$ in \eref{eq:vowf} is the valley-dependent envelope function. As we will see in the following (shown originally in \ocite{Friesen07}), in the absence of interface steps and magnetic field, the envelope function of the complete ground state at the leading order contains only the orbital ground state, and it is independent on the valley state. In this special and ideal case, the valley and orbital indices are good quantum numbers. 

However, particularly in the presence of interface steps, the envelope functions $\Psi_{xyz}^{j}$ will contain not only the orbital ground state but also the orbital excited states. Furthermore, the dependency of the envelope functions on the valley state also becomes more important. The Schr\"odinger equation governing  the valley-dependent envelope functions for  strained silicon is  \cite{Friesen07},
\begin{align}
\label{eq:VDenv}
\sum_{j=\pm z}a_je^{ik_jz}\left\{H_c+V_v(r)-E\right\}\Psi_{xyz}^{j}=0.
\end{align}
Here, $V_v(r)$ is the valley coupling parameter  \cite{Friesen07}  that  vanishes everywhere except at the Si/barrier interface, at $r=r_\mathrm{int}$, and from which we can deduce the valley splitting.
The term $H_c$ in \eref{eq:VDenv}  describes the electron confinement. Assuming a SiGe/Si/SiGe quantum dot with ideal Si/SiGe interface, in the absence of a magnetic field, we can write $H_c=H_0$ with,
\begin{align}
\label{eq:H_xyz_B0}
H_{0}=&\frac{p_x^2}{2m_t}+\frac{1}{2}m_t\w_x^2x^2 +\frac{p_y^2}{2m_t}+\frac{1}{2}m_t\w_y^2y^2\nonumber\\&+\frac{p_z^2}{2m_l}-eF_zz+ U(z),
\end{align}
where $m_t=0.19\:m_e$ and $m_l=0.98\:m_e$ are the transverse and longitudinal effective mass, and $\w_{x}=\hbar/m_tx_0^2$ and $\w_{y}=\hbar/m_ty_0^2$ are the confinement frequencies along $\hat{x}$ and $\hat{y}$. Following \ocite{Hosseinkhani20}, the out-of-plane potential profile for a SiGe/Si/SiGe reads 
\begin{align}
\label{eq:Uz}
U(z)=U_0\theta(-z-d_t) + U_0\theta(z) + U_{\infty}\theta(z-d_b),
\end{align}
where $U_0=150$ meV is the energy offset between the minima of the conduction band in Si and Si$_{1-x}$Ge$_x$ with $x=0.3$,  $d_t$ is the thickness of the silicon layer (located between $-d_t\leq z\leq 0$) and $d_b$ is the thickness of the upper SiGe barrier. 

We now perform a perturbation theory in the valley coupling $V_v$ in \eref{eq:VDenv}. At the zeroth order, we ignore the valley coupling, and see that because of the fast oscillations due to the exponential factor, $e^{ik_jz}$, the contribution from the two valleys become (nearly) decoupled. As such, from \eref{eq:VDenv} we arrive to the  Schr\"odinger equation below for the valley-independent envelope function,
\begin{align}\label{eq:schH0}
    H_0\psi_{xyz}=\e\psi_{xyz}.
\end{align}
This equation has been the starting point for many works concerned with studying the valley splitting, including but not limited to Refs.~\cite{Koiller09,Koiller11,Culcer12,Hosseinkhani20}. 

Given the confinement Hamiltonian $H_0$, we can write for the ground state envelope function $\psi_{xyz,0}=\psi_{x,0}\psi_{y,0}\psi_{z,0}$. Here, the in-plane envelope functions are trivially given by the well known harmonic-oscillator wave functions. The out-of-plane envelope functions $\psi_{z,n}$ are discussed in detail in \ocite{Hosseinkhani20}. In particular, for the ground state we can write,
\begin{equation}
\label{eq:psi_0z}
  \psi_{z,0}(\tz) \simeq\\
   \frac{z_0^{-1/2}}{\Ai'(-r_0)} 
  \begin{cases}
      \Ai(-\te_{z,0})e^{-\frac{\Ai'(-\te_{z,0})}{\Ai(-\te_{z,0})}\tz}\: , & \tz > 0 \, \\
   \Ai(-\tz-\te_{z,0})\: . & \tz \leq 0\, 
  \end{cases}
\end{equation}
while the (normalized) ground state energy reads,
\begin{align}
\label{eq:E_0z}
\te_{z,0} \simeq r_0 -\tilde{U}_0^{-1/2}.
\end{align} 
Here Ai is the Airy function, Ai$'$ its first derivative, and $-r_0\simeq-2.3381$ its  smallest root (in absolute value). We also used normalized position, $\tz=z/z_0$, energy, $\te_{z,0}=\e_{z,0}/\e_0$, and potential $\tu_0=U_0/\e_0$ for which the length and energy scales read, respectively,
\begin{align}
    z_0 &= \left[\frac{\hbar^2}{2m_leF_z}\right]^{1/3},&
    \e_0 = \frac{\hbar^2}{2m_lz_0^2}.
    \label{eq:e0}
\end{align}
For later use, we note that by substituting \eref{eq:E_0z} into (\ref{eq:psi_0z}) and expanding $\te_{z,0}$ around $r_0$, we find for the ground state at the interface position at the leading order,
\begin{align}\label{eq:psi_zAT0}
    \psi_{z,0}(z=0)\simeq\sqrt{eF_z/U_0}.
\end{align}
As noted in \ocite{Hosseinkhani20}, we emphasize that \esref{eq:psi_0z} and (\ref{eq:E_0z}), and therefore \eref{eq:psi_zAT0} as well, are valid provided that the amplitude of the envelope function at the lower barrier/Si interface is negligible. Assuming the thickness of the Si layer is $d_t=10$ nm, $F_z\gtrsim 2$ MV/m validates the assumption of neglecting the lower interface. Throughout this paper, we consider circular SiGe/Si/SiGe quantum dots where the relevant value for the electric field is typically $F_z=15$ MV/m \cite{Culcer12,Koiller09}. As such, neglecting the lower interface is well justified.
In \ocite{Hosseinkhani20}
it is also discussed in detail how to find the excited states of the out-of-plane electron motion $\psi_{z,n\geq 1}$ and $E_{z,n\geq 1}$. Knowing the excited states is essential to carry on with our analysis, and we take them as given quantities in this paper.
\begin{figure}[t!]
\begin{center}
\includegraphics[width=0.45\textwidth]{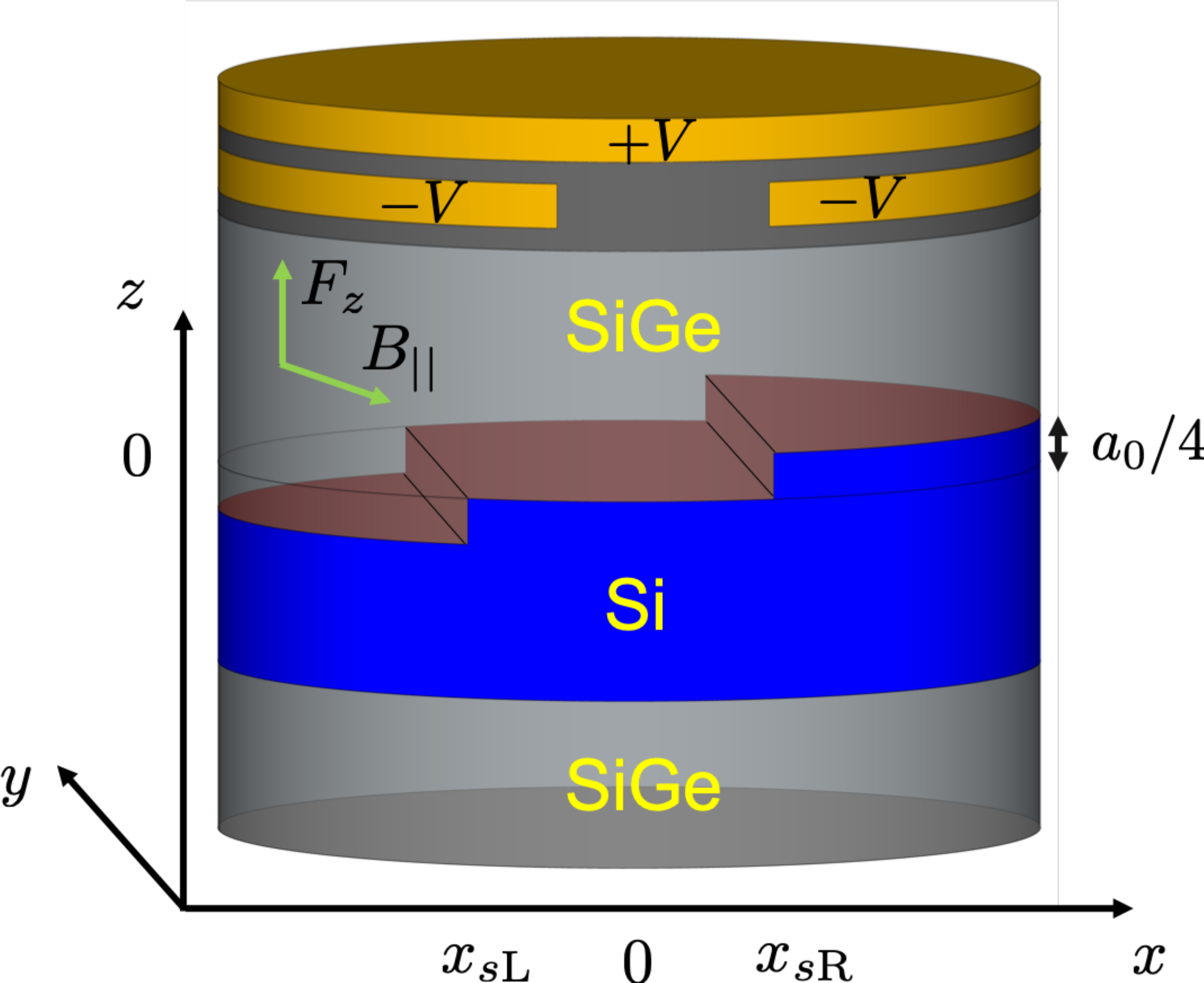}
\end{center}
\caption{Schematic of a quantum dot with stair-like disordered interface. The  top gates with applied voltages $\pm V$ are used to trap and confine a single
electron in the silicon layer. The pink area marks the upper Si/SiGe interface. The single-layer atomic steps have the width $a_0/4$ where $a_0=0.543$ nm denotes the lattice constant. }  \label{fig:dis_dot}
\end{figure} 

Let us now consider the general case where we take into account the valley-coupling parameter $V_v(r)$, and allow the presence of interface steps and an in-plane magnetic field, $\textbf{B}_{||}=(B_{x},B_{y},0)=B(\cos\phi_B,\sin\phi_B,0)$. For simplicity, throughout this paper we assume that the interface steps are all parallel to the $\hat{y}$-axis. We further assume there are two single-layer interface steps located at the left and right side of the dot center, at $x_{s\mathrm{L}}\leq 0$ and $x_{s\mathrm{R}}\geq 0$, as depicted in Fig.~\ref{fig:dis_dot}. As we show in detail in Appendix~\ref{app:PC}, within the first order in the perturbation, the valley-dependent envelope function from \eref{eq:VDenv} for the ground ($q=0$) and the first excited ($q=1$) valley-orbital states read,
\begin{align}
    \label{eq:Psi_t_p} %
    \Psi_{xyz}^{\pm z,(q)}= \psi_{xyz,0} + \psi_{||} +  \psi_\mathrm{st}^{\pm z,(q)}.
\end{align}

\begin{figure*}[t!]
\begin{center}
\includegraphics[width=1\textwidth]{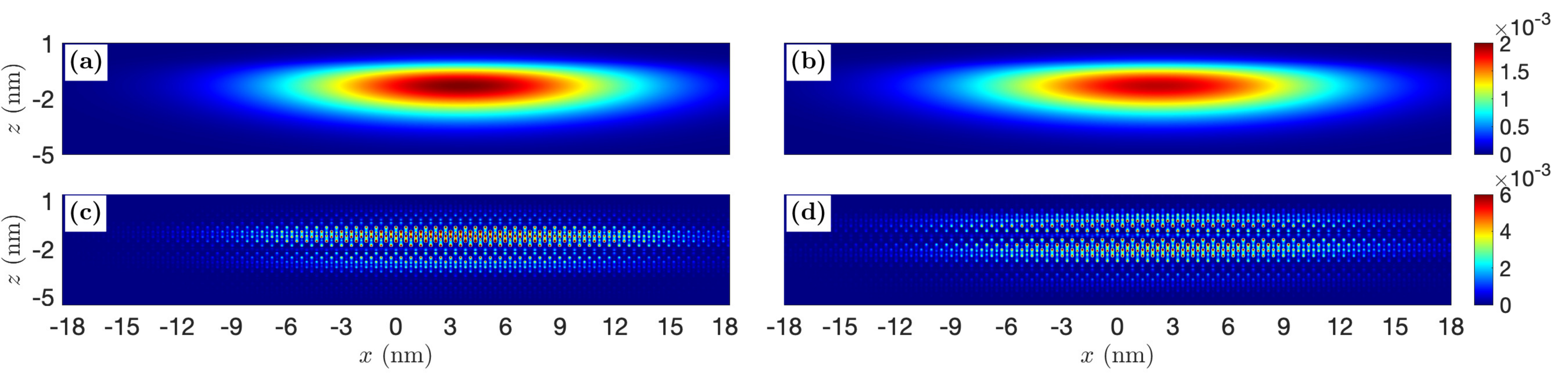}
\end{center}
\caption{[(a) and (b)]  Amplitude squared of the envelope function for the ground (a) and excited (b) valley-orbital state in the $x-z$ plane, $|\Psi_{xyz,0}^{+z,(q=0)}(x,y=0,z)|^2$ and $|\Psi_{xyz,0}^{+z,(q=1)}(x,y=0,z)|^2$. [(c) and (d)]  Probability density of the ground (c) and excited (d) valley-orbital state, $P_0=\langle\nu^{(q=0)}|\nu^{(q=0)}\rangle$   and   $P_1=\langle\nu^{(q=1)}|\nu^{(q=1)}\rangle$ in the $x-z$ plane. The results shown in all panels have the unit of $(1/\mathrm{nm}^3)$.  Here we set $B=0$, $x_{s\mathrm{L}}=-x_0$,  $x_{s\mathrm{L}}=0.265x_0$, $\hbar\omega_x =3.9$ meV (that gives $x_0=\sqrt{\hbar/m_t\omega_x}\simeq 10.14$ nm) and $F_z=15$ MV/m (which gives rise to $z_0\simeq 1.4$ nm.)}  \label{fig:DP_PD}
\end{figure*} 
Here, for the correction to the envelope function due to the presence of an in-plane magnetic field $\textbf{B}_{||}$, we find the similar result as presented in \ocite{Hosseinkhani20},
\begin{align}\label{eq:psi_par}
    \psi_{||}=&-iB_x\psi_{x,0}\psi_{y,1}\sum_{n=1}\alpha_n\psi_{z,n}\\
&+iB_y\psi_{x,1}\psi_{y,0}\sum_{n=1}\beta_n\psi_{z,n}
-B_xB_y\eta\psi_{x,1}\psi_{y,1}\psi_{z,0}\:.\nonumber
\end{align}
The valley-dependent correction to the envelope function due to the presence of interface steps and the valley-coupling reads,
\begin{align}\label{eq:psi_st}
    \psi_\mathrm{st}^{\pm z,(q)}=\psi_{y,0} \sum_{(m,n)\neq(0,0)} c_{m,n}^{ \pm z,(q)}\psi_{x,m}\psi_{z,n},
\end{align}
where the coefficients are related via the time-reversal symmetry relation, $c_{m,n}^{ +z,(q)}=[c_{m,n}^{-z,(q)}]^*$.

The exact definitions of the perturbative coefficients used when writing $\psi_{||}^{+z}$ and $\psi_\mathrm{st}^{\pm z, (q)}$  and  for their numerical calculation are given in Appendix \ref{app:PC}. In Tables~\ref{tab:par_alpha} and \ref{tab:par_cmn} of Appendix~\ref{app:PC} we show an example for the obtained values for the perturbative coefficients in a disordered quantum dot with realistic parameters. We note here that, as shown in \ocite{Hosseinkhani20}, in a Si/SiGe quantum dot and at the realistic value of $F_z=15$ MV/m, there are only 3 out-of-plane excited states with $\e_{n,z}<U_0$ (the excited states energies are shown in Table~\ref{tab:par_alpha}.) Therefore, in the summation over $n$ in \esref{eq:psi_par} and (\ref{eq:psi_st}), we can set $n_\mathrm{max}=3$ as the cutoff. Moreover, we find that by taking up to 4 in-plane excited states, the coefficients $c_{m,n}^{ +z,(q)}$ in \eref{eq:psi_st} substantially decay. Therefore, we set $m_\mathrm{max}=4$ as the cutoff.

Having found the valley-dependent envelope function at the first order in the perturbation, we also find for the valley splitting,
\begin{align}\label{eq:VS_dis}
    E_{vs}^\mathrm{dis}=E^{(q=0)}-E^{(q=1)}=2|\Delta_1^\mathrm{dis}|,
\end{align}
in which we have,
\begin{align}\label{eq:Delta_1_dis}
    \Delta_1^\mathrm{dis} = V_v\int e^{-2ik_0z} \mathcal{S}_\mathrm{int}(x,z)\psi_{xyz,0}^2 d^3r,
\end{align}
and, therefore, the valley phase reads, $\phi_v=\mathrm{arg}[\Delta_1^\mathrm{dis}]$. The parameter $V_v$ is given by \eref{eq:Vv} in Appendix~\ref{app:PC} where we model it as a function of the out-of-plane electric field $F_z$, the offset potential $U_0$ and the details of the periodic part of the Bloch function $u_{\pm z}(r)$. The interface function $\mathcal{S}_\mathrm{int}(x,z)$ vanishes everywhere except at the interface and it is given by \eref{eq:Sint} for a disordered quantum dot within our model.

We can now also obtain the full valley-orbital wave function from the effective mass theory. Let us first define the pure valley states,
\begin{align}\label{eq:pmZ}
|\pm z^{(q)}\rangle=&e^{\pm ik_0z}u_{\pm z}(r)\Psi_{xyz}^{\pm z,(q)},
\end{align}
where $q=0,1$.
Using these as well as \eref{eq:vowf}, the valley-orbital wave function, up to a global phase factor, can be written as,
\begin{align}\label{eq:v0}
    |\nu^{(q=0)}\rangle=\frac{1}{\sqrt{2}}\left\{|+z^{(0)}\rangle-e^{-i\phi_v}|-z^{(0)}\rangle\right\},\\\label{eq:v1}
    |\nu^{(q=1)}\rangle=\frac{1}{\sqrt{2}}\left\{|+z^{(1)}\rangle+e^{-i\phi_v}|-z^{(1)}\rangle\right\}. 
\end{align}

 In Fig.~\ref{fig:DP_PD}[(a) and (b)], we show the amplitude squared of the ground and first excited envelope function, $|\Psi_{xyz}^{+z,(q=0)}|^2$ and $|\Psi_{xyz}^{+z,(q=1)}|^2$, in the $x-z$ plane for a disordered quantum dot. 
 The coefficients $c_{m,n}^{+z, (q=0)}$ and $c_{m,n}^{+z, (q=1)}$ obtained for this disordered quantum dot are shown in Table~\ref{tab:par_cmn} in Appendix~\ref{app:PC}. As can been seen from Fig.~\ref{fig:DP_PD}, due to the presence of the interface steps, the two envelope functions lack mirror symmetry along $\hat{x}$. Furthermore, the envelope functions of the ground and excited state are different from each other. This, in turn, causes the intra-valley and inter-valley dipole moment along $\hat{x}$ to become non-zero. We study this in detail in the following subsection. We have also shown in  Fig.~\ref{fig:DP_PD}[(c) and (d)] the  probability density of the ground  and first excited valley-orbital state in the $x-z$ plane. Here we used \ocite{Koiller11} in finding the periodic parts of the Bloch function, $u_\pm(z)$. All wave functions in Fig.~\ref{fig:DP_PD} are shown with their actual aspect ratio, and the fact that the out-of-plane confinement is much stronger than the in-plane conferment is clearly visible. The assumed locations for the interface steps for Fig.~\ref{fig:DP_PD} are $x_{s\mathrm{L}}=-x_0$ and $x_{s\mathrm{L}}=0.265x_0$. We note that by using this specific choice for the location of interface steps, we were able to fit some experimental data for the spin qubit relaxation time originally presented by \cite{Borjans19}, as we discuss in Sec.~\ref{sec:dis} [see Figure~\ref{fig:T_1_the_vs_exp}.]

\subsection{Dipole matrix elements}\label{subsec:dip}
We now turn to consider the dipole matrix elements between the two low-lying valley-orbital states. Our objective here is to study how the interface roughness, the dot lateral size (which sets the in-plane orbital splitting) and the electromagnetic field influence the dipole matrix elements. Using \esref{eq:v0} and (\ref{eq:v1}), for any operator \textit{O} we can write, 
 \begin{align}\label{eq:O_qq'}
&\langle \nu^{(1)}|\textit{O}|\nu^{(0)}\rangle=
\frac{1}{2}\Big\{\langle +z^{(1)}|\textit{O}|\! +\! z^{(0)}\rangle-\langle \! -\! z^{(1)}|\textit{O}|\!-\! z^{(0)}\rangle \nonumber\\
&- e^{-i\phi_v}\langle +z^{(1)}|\textit{O}|\!-\! z^{(0)}\rangle+e^{i\phi_v}\langle \!-\! z^{(1)}|\textit{O}|\!+\! z^{(0)}\rangle\Big\},
\end{align}
and
\begin{align}\label{eq:O_qq}
 &\langle \nu^{(q)}|\textit{O}|\nu^{(q)}\rangle=
 \frac{1}{2}\Big\{\langle +z^{(q)}|\textit{O}|\! +\! z^{(q)}\rangle+\langle \! -\! z^{(q)}|\textit{O}|\!-\! z^{(q)}\rangle \nonumber\\
 &+ e^{-i\phi_v}\langle +z^{(q)}|\textit{O}|\! -\! z^{(q)}\rangle+e^{i\phi_v}\langle -z^{(q)}|\textit{O}|\!+\! z^{(q)}\rangle\Big\}.
 \end{align}

We note here that as long as the dipole matrix elements are concerned, the last two terms in both of the above equations are negligible; this is due to the fast-oscillating factor of $e^{\pm 2ik_0z}$ inside the integrand of those terms.  We then immediately see that if one starts from a valley-independent envelope function theory, \eref{eq:schH0}, the inter-valley dipole matrix element, \eref{eq:O_qq'}, vanishes even in the presence of interface steps (it is straightforward to verify $\langle +z|\textit{r}|\!+\! z\rangle = \langle -z|\textit{r}|\!-\! z\rangle$ in this case; note that the q-index is naturally not relevant within the valley-independent envelope function theory.)

However, using the valley-dependent envelope function theory that we developed earlier in this section, we are now able to calculate both the inter-valley and intra-valley dipole matrix elements. The $x$ dipole matrix elements read (see Appendix \ref{app:dipole} for details),
\begin{align}\label{eq:ME_x_nd}
    \langle\nu^{(q)}|\hat{x}|\nu^{(q')}\rangle & =  \frac{x_0'}{2}\mathrm{Im}\left[c_{1,0}^{-z,(q)}+c_{1,0}^{+z,(q')}\right]\\
\label{eq:ME_x_d}
 \langle\nu^{(q)}|\hat{x}|\nu^{(q)}\rangle & =  x_0'\mathrm{Re}\left[c_{1,0}^{+z,(q)}\right],
\end{align}
while the $z$ dipole matrix elements at the leading order become
\begin{align}\label{eq:ME_z_nd}
    \langle\nu^{(q)}|\hat{z}|\nu^{(q')}\rangle &= 
    \sum_{n=1}^{3}\mathrm{Im}\left[c_{0,n}^{-z,(q)} +c_{0,n}^{+z,(q')}\right]s_{n},\\\label{eq:ME_z_d}
\langle\nu^{(q)}|\hat{z}|\nu^{(q)}\rangle &=  s_0 +2\sum_{n=1}^{3}\Re\left[c_{0,n}^{+z, (q)}\right]s_n,
\end{align}
where in the above relations $q\neq q'$ and we defined $s_{n}=\int_{-\infty}^{+\infty} z \psi_{z,0}\psi_{z,n}dz$.
Given that the $z$ inter-valley dipole matrix element is entirely originating from the coupling to the out-of-plane excited states, we conclude that generally this matrix element is very small due to the strong out-of-plane confinement in quantum dots. On the other hand, we see that the dominant contribution to \eref{eq:ME_z_d} only depends on $\psi_{z,0}$. This indicates that the $z$ intra-valley dipole matrix element is only slightly influenced by interface roughness.
\begin{figure}[b!]
\begin{center}
\includegraphics[width=0.48\textwidth]{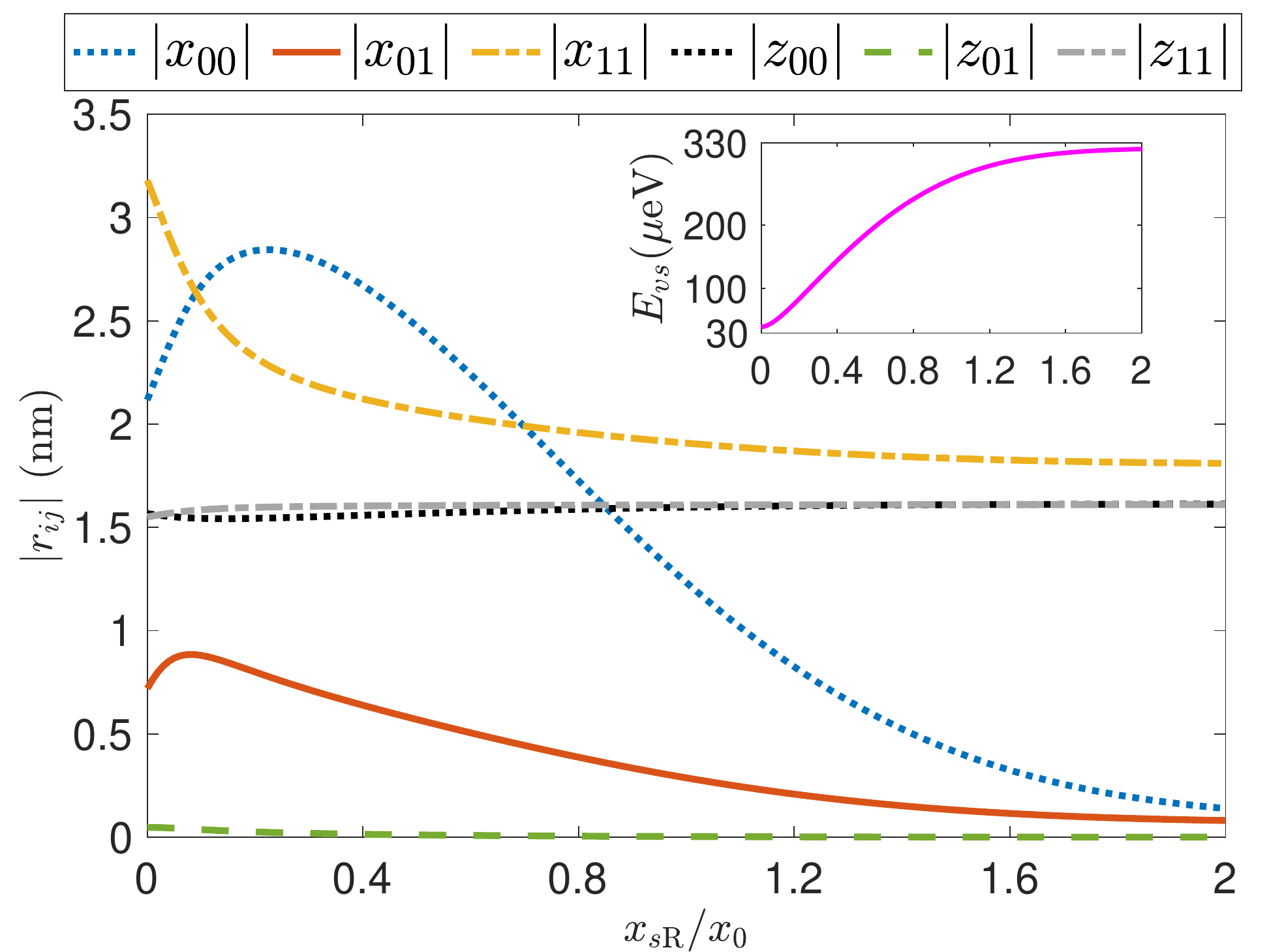}
\end{center}
\caption{The inter-valley and intra-valley dipole matrix elements defined by $r_{ij}=\langle \nu^{(q=i)}|\textit{r}|\nu^{(q=j)}\rangle$ as a function of the position of the interface steps located at $x_{s\mathrm{R}}$. Inset: Valley splitting as a function of $x_{s\mathrm{R}}$. Here we set all the other parameters to be the same as given in the caption of Fig.~\ref{fig:DP_PD}.} \label{fig:r_ij}
\end{figure}

In Fig.~\ref{fig:r_ij}, we show the obtained dipole matrix elements from \esref{eq:ME_x_nd} to (\ref{eq:ME_z_d}) as a function of the position of the step located at $x_{s\mathrm{R}}$ whereas we fixed $x_{s\mathrm{L}}=-x_0$.  As expected, we observe that the $z$ intra-valley dipole moment remains nearly constant when changing the step location while the $z$ inter-valley dipole moment is always very small. We observe that for $x_{s\mathrm{R}}\gtrsim x_0$ the dipole moment $x_{11}$ starts to saturate to a finite value whereas the dipole moments $x_{00}$ and $x_{01}$ decrease much faster and nearly vanish by placing $x_{s\mathrm{R}}$ far away from the quantum dot center. The saturation observed in the behavior of $x_{11}$ can be attributed to the presence of the other step at $x_{s\mathrm{L}}$. In the presence of the stair-like interface steps as considered in our model, see Fig.~\ref{fig:dis_dot}, the silicon quantum well is thicker at the right side of the quantum dot. As such, it is energetically favorable for the electron wave function to shift towards the right side of the quantum dot, as observed in Fig.~\ref{fig:DP_PD}. However, since the valley-orbital excited state $q=1$ has higher energy than the ground state, the wave function has further spatial spread within the quantum well compared to the ground state, and particularly the excited state envelope function has a larger amplitude at $x_{s\mathrm{L}}$. In the inset of Fig.~\ref{fig:r_ij}, we show the valley splitting \eref{eq:VS_dis} as a function of the position of the interface step located at $x_{s\mathrm{R}}$. The suppression of the valley splitting due to the presence of interface steps is discussed in detail in \ocite{Hosseinkhani20}.

\section{interface-induced spin-orbit interaction}\label{sec:SOI}
The interface inversion asymmetry  gives rise to a Rashba-like, $H_R$, and a Dresselhaus-like, $H_D$, spin-orbit interaction \cite{Golub04}. For an ideally flat quantum dot with the upper Si/SiGe interface at $z_i=0$ we have,
\begin{align}\label{eq:H_R_I}
H_R^\mathrm{ideal}=\gamma_R(p_y\sigma_x-p_x\sigma_y)\delta(z),\\\label{eq:H_D_I}
H_D^\mathrm{ideal}=\gamma_D(p_x\sigma_x-p_y\sigma_y)\delta(z),
\end{align}
where $\sigma_{x(y)}$ are Pauli matrices and $\gamma_{R(D)}$ are $2\times2$ matrices in the two-dimensional valley space. Here the lower interface is neglected since due to the electric field, the amplitude of the wave function is negligible, as discussed in Sec.~\ref{subsec:wave}.  Note that the presence of the delta function in the above equations is to ensure that the spin-orbit interaction quickly vanishes away from the interface. This has been the justification in some works to integrate over the out-of-plane degree of freedom, and consider the spin-orbit interaction only for 2D electrons. In this case, the 2D spin-orbit coefficients become a function of the applied electric field and this dependence is found to be linear for both the Rashba and Dresselhaus terms \cite{Prada11,Ruskov18}.

Here we aim to take into account the influence of interface roughness on the spin-orbit interaction. As such, we keep considering the spin-orbit interaction for 3D electrons. In the presence of the interface steps, the 3D Rashba and Dresselhaus-like spin-orbit terms, \esref{eq:H_R_I} and (\ref{eq:H_D_I}), have to be generalized to include the interface function $\mathcal{S}_\mathrm{int}(x,z)$, given by \eref{eq:Sint} in our model, instead of the delta function. We note that in general the momentum operator may not commute with the interface function, $[p_{x(y)} , \mathcal{S}_\mathrm{int}(\mathbf{r})]\neq 0$. Therefore, symmetrization of $p_{x(y)}$ and $\mathcal{S}_\mathrm{int}$ is required. Such symmetrization is analogous to the one performed in systems where, due to the local disorder, the coefficients involve in spin-orbit interaction are dependent on the in-plane coordinates \cite{Glazov10,Porubaev14}. Moreover, we note that due to the crystal symmetry of silicon, a vertical shift of the interface location due to a single atomic step, $z_i\rightarrow z_i + a_0/4$ is equivalent to an in-plane rotation by $\pi/2$. This, in turn, indicates that the Dresselhaus term must change sign under such transformation whereas the Rashba term remains the same \cite{Ferdous18,FerdousPRB18,Jock18}. We therefore write for the symmetrized spin-orbit interaction in the presence of interface steps,
\begin{align}\label{eq:H_R}
    H_R&=\frac{1}{2}\gamma_R\left\{ p_y\sigma_x-p_x\sigma_y,\mathcal{S}_\mathrm{int}(x,z)\right\},\\\label{eq:H_D}
    H_D&=\frac{1}{2}\gamma_D\cos\left(\frac{4\pi z}{a_0}\right)\left\{p_x\sigma_x-p_y\sigma_y,\mathcal{S}_\mathrm{int}(x,z)\right\}.
\end{align}
Here $\{B,C\}=BC+CB$ is the anti-commutator and the factor of $\cos(4\pi z/a_0)$ in the Dresselhaus term ensures the necessary sign change caused by a single-layer interface step.

We note that the above forms for the spin-orbit interaction is alined with the observation that the spin-orbit interaction is sample-dependent, as the matrix elements of the Rashba and Dresselhaus terms, $\langle\nu^{(q)}|H_{R(D)}|\nu^{(q')}\rangle$, depends on the form of the interface function $\mathcal{S}_\mathrm{int}$ that can vary across different samples. To proceed, we take the coefficients to be $\gamma_R=A_0 \gamma_R^{0}$ and $\gamma_D=A_0 \gamma_D^{0}$ for which we consider $A_0$ as a fitting parameter and we use the theoretical analysis of \ocite{Prada11} to extract the coefficients $\gamma_R^{0}$ and $\gamma_D^{0}$ (that are defined for 3D electrons). This reference applies  atomistic calculations to an ideally flat Si$_{0.7}$Ge$_{0.3}$/Si/Si$_{0.7}$Ge$_{0.3}$ heterostructure and it finds,
\begin{align}\label{eq:alpha_RD}
    \gamma_{R(D)}^\mathrm{2D}=\gamma_{R(D)}^{0}\int\psi_{z,0}^2\delta(z)dz=\alpha_{R(D)}F_z,
\end{align}

Using \eref{eq:psi_zAT0} for the amplitude of the envelope function at the interface together with the above equations, we find the Rashba and Dresselhaus coefficients of the spin-orbit interaction presented in Table~\ref{tab:soi_coef}. We note that the fitting parameter $A_0$ does not need to be the same in  the Rashba and Dresselhaus terms. However, here we assume this for simplicity, and we show in Sec.~\ref{sec:dis} that this simplified model can successfully match experimental measurements.
Furthermore, we note that similar to the theoretical prediction in \ocite{Prada11}, experimental measurements also indicate that the Dresselhaus term is much stronger than the Rashba term, $\gamma_R\ll\gamma_D$ \cite{Ferdous18,Tanttu19}.
\begin{table}[t!]
\begin{center}
\caption{\label{tab:soi_coef} The coefficients for the inter-valley and intra-valley spin-orbit interaction. The terms $\alpha_R$ and $\alpha_D$ are introduced in \eqref{eq:alpha_RD} and their valued are reported from \ocite{Prada11}.}
\begin{tabular}{c  c  c } \hline \hline
  & inter-valley   & intra-valley  \\ \hline 
$\alpha_R\,(\mathrm{e}\cdot \mathrm{nm}^2/\hbar)$ & $1.5\times 10^{-5}$ & $0.7\times 10^{-5}$ \\ 
$\alpha_D\,(\mathrm{e}\cdot \mathrm{nm}^2/\hbar)$ & $97.8\times 10^{-5}$ & $30.6\times 10^{-5}$ \\ 
$\gamma_R^0\,(\mu\mathrm{eV}\cdot\mathrm{nm}^2/\hbar)$ & $2.25$ & $1.05$ \\ 
$\gamma_D^0\,(\mu\mathrm{eV}\cdot \mathrm{nm}^2/\hbar)$ & $146.70$ & $45.90$ \\\hline\hline
\end{tabular}
\end{center}
\end{table}

\section{Spin-Qubit levels}\label{sec:SQL}

The logical states of an ideal spin qubit should contain only the spin-down state at the qubit ground state and the spin-up state at the qubit excited state \cite{LossDiVincenzo98}. However, due to the valley and orbital excitations and the spin-orbit interaction, the qubit logical ground state acquires another component including the spin-up state and, likewise, the logical excited state acquires a component including the spin-down state. This, in turn, enables qubit relaxation due to the spin-conserving electron-phonon interaction as well as the Johnson and $1/f$ charge noise. 

Here we consider modifications to the spin-qubit levels due to the spin-valley mixing (SVM) as well as the spin-orbit mixing (SOM). In particular, it has already been shown that when the Zeeman energy $E_z=g\mu_BB$ and valley splittings coincide, the spin-valley mixing gives rise to a hotspot at which the qubit relaxation time is substantially reduced. However, at higher magnetic fields where the Zeeman energy becomes sufficiently larger than the valley splitting, the dominating contribution to the qubit relaxation turns to be due to the spin-orbit mixing. In this section, we use our findings for the valley-dependent envelope functions as well the interface-induced spin-orbit interaction to calculate the corrections to the spin-qubit levels due to both SVM and SOM. This, in turn,  enables us to study the qubit relaxation as a function of interface roughness as well as the electromagnetic fields.   

\subsection{Qubit-level modification due to SVM}\label{sec:SQL_SVM}
Here we assume that the Zeeman energy is much smaller than the orbital splitting so that it is sufficient for us to only consider the two low-lying valley-orbital states given by \esref{eq:v0} and (\ref{eq:v1}). We then consider only the following unperturbed states,
\begin{align}
\label{eq:4b}
  |1\rangle &= |\nu^{(q=0)},\downarrow\rangle ,\quad |2\rangle = |\nu^{(q=0)},\uparrow\rangle , \nonumber \\
  |3\rangle &= |\nu^{(q=1)},\downarrow\rangle ,\quad |4\rangle = |\nu^{(q=1)},\uparrow\rangle.
\end{align}
The spin-orbit interaction couples the above states to each other and, therefore, it modifies the qubit levels. To find the modified states, often in the literature a 2D spin-orbit interaction is employed, and it is then argued that $p_{x}=(im_t/\hbar)[H_c,x]$ (and similarly for $p_y$)\cite{Yang13, Huang14}. Based on this, one arrives for inter-valley matrix elements $\langle\nu^{(q=1)}|p_x|\nu^{(q=0)}\rangle=(im_tE_{vs}/\hbar)x_{10}$, where the dipole moment is treated as a free parameter. We stress here that this approach is valid only in the absence of a magnetic field. Furthermore, as noted in the previous section, in the presence of interface roughness, it is appropriate to use \esref{eq:H_R} and (\ref{eq:H_D}) for the spin-orbit interaction in which the presence of the interface function $\mathcal{S}_\mathrm{int}$ prevents the use the above commutation relation. Using the valley-dependent envelope function \eref{eq:Psi_t_p}, we are now able to calculate the matrix elements of the spin-orbit interaction, $\Delta_{ij}=\langle i|H_R+H_D|j\rangle$, as a function of the interface roughness and the electromagnetic fields. We find for the inter-valley coupling (see Appendix~\ref{app:ME_SOI} for details),
\begin{align}\label{eq:D_32}
\Delta_{32}&=-i\gamma_R\cos\phi_Bg_c -i\gamma_D\sin\phi_Bg_c'\nonumber\\ &+\gamma_R\sin\phi_v\left(B_yf_\beta\cos\phi_B-B_xf_\alpha\sin\phi_B\right)\nonumber\\
   &+\gamma_D\sin\phi_v\left(B_yf_\beta'\sin\phi_B-B_xf_\alpha'\cos\phi_B\right),
\end{align}
and $\Delta_{41}=-\Delta_{32}$. For the intra-valley coupling we find,
\begin{align}\label{eq:D_21}
    \Delta_{21}=i&\left(1+\cos\phi_v\right)\big[\gamma_R\left(B_yf_\beta\cos\phi_B-B_xf_\alpha\sin\phi_B\right)\nonumber\\
    &+\gamma_D\left(B_yf_\beta'\sin\phi_B-B_xf_\alpha'\cos\phi_B\right)\big],
\end{align}
and $\Delta_{43}=\Delta_{21}$. We recall here that coefficients of the spin-obit interaction, $\gamma_R$ and $\gamma_D$, are valley-dependent, see Table~\ref{tab:soi_coef}.

The  definitions of  $f_{\alpha(\beta)}$, $f_{\alpha(\beta)}'$, $g_c$, and $g_c'$ are given in Appendix~\ref{app:ME_SOI}. We note that all of these terms are real quantities, the terms with (without) the prime are due to the Dresselhaus (Rashba) interaction, and while $f_{\alpha(\beta)}$ and $f_{\alpha(\beta)}'$ originate from the magnetic-field induced coupling between the (unperturbed) orbital ground state to the out-of-plane excited states, $g_c$ and $g_c'$ originate from the presence of interface steps and the valley coupling. Therefore,  $g_c$ and $g_c'$ involve  coupling  of  the  orbital ground  state  to  both  in-plane  and  out-of-plane  excited states.
\begin{figure}[t!]
\begin{center}
\includegraphics[width=.46\textwidth]{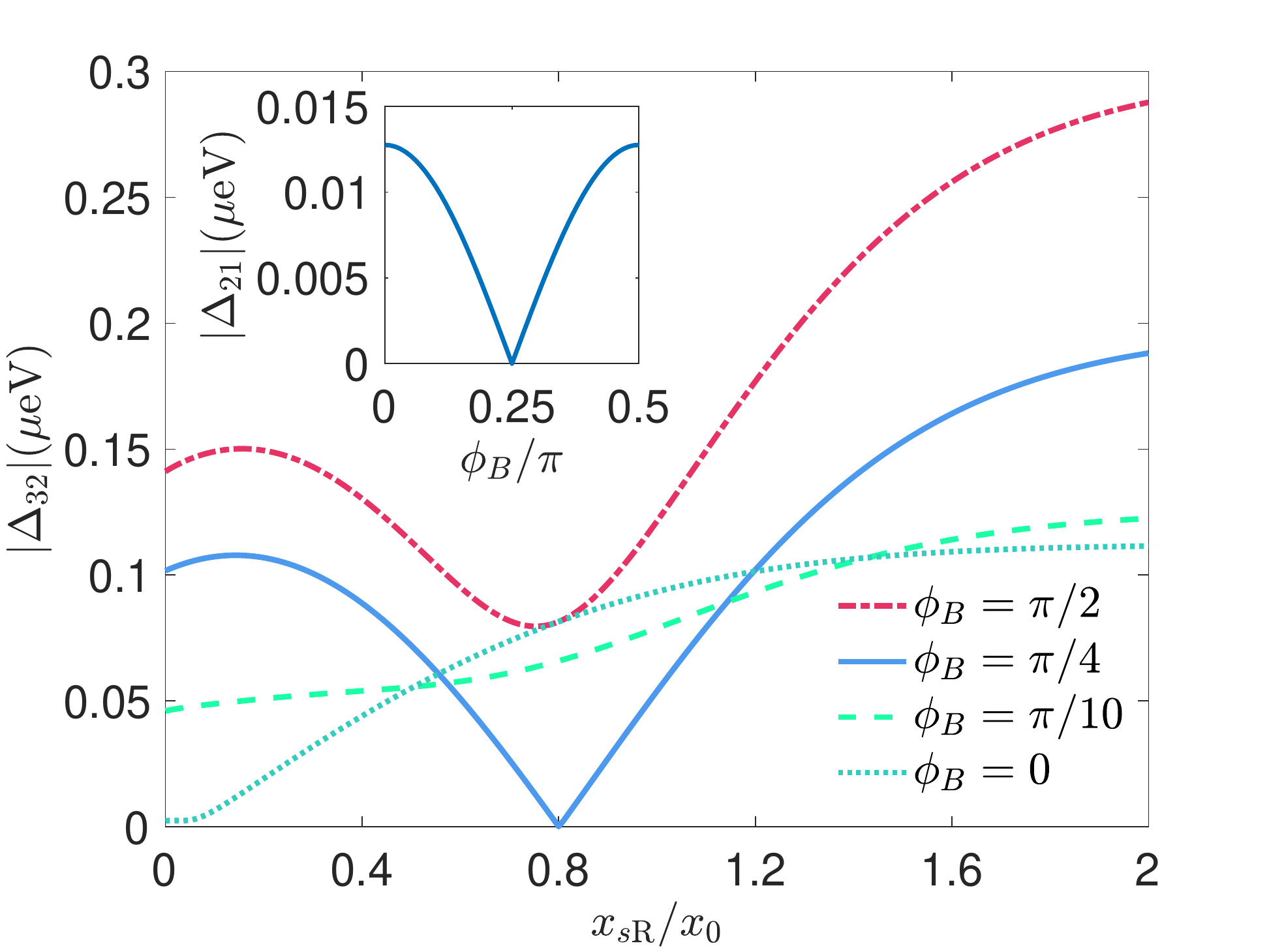}
\end{center}
\caption{The inter-valley spin-valley coupling $\Delta_{32}(\phi_B)$ for various directions of the magnetic field as a function of the atomic step at $x_{s\mathrm{R}}$. Here we fixed $x_{s\mathrm{L}}=-x_0$. Inset: Intra-valley spin-valley coupling $\Delta_{21}(\phi_B)$ as a function of the direction of the magnetic field. Here we set $x_{s\mathrm{L}}=-x_0$ and $x_{s\mathrm{R}}=0.265x_0$. For both panels we assumed $B=1$ T. All other parameters are the same as used in Fig.~\ref{fig:T_1_the_vs_exp}, in particular $A_0=5.42$. }  \label{fig:D_32}
\end{figure}

Given \esref{eq:D_32} and (\ref{eq:D_21}), we realize that the inter-valley coupling $\Delta_{32}$ is in general a complex quantity whereas the intra-valley coupling is purely imaginary. However, in a circular quantum dot when the direction of the magnetic field is $\phi_B=\pi/4$, one finds $\Delta_{21}=0$ while $\Delta_{32} \propto i\gamma_Rg_c + i\gamma_Dg_c'$ becomes purely imaginary. Remarkably, in this case, and for certain configurations for the interface steps, the inter-valley coupling can vanish. This happens when $g_c'=-(\gamma_R/\gamma_D)g_c$ which is possible due to the sign change in the Dresselhaus term caused by single-layer atomic steps.  As we show later in Sec.~\ref{sec:dis}, extinction of the spin-valley coupling has a profound effect on the qubit relaxation as it  completely removes the valley-induced qubit decay.  

In Fig.~\ref{fig:D_32}, we show the inter-valley coupling for various directions of the magnetic field (at $B=1$ T) as a function of the location of the step at $x_{s\mathrm{R}}$. The vanishing of $\Delta_{32}(\phi_B=\pi/4)$ at a certain position for the interface step is clear in the figure. Moreover, we observe that the $\Delta_{32}$ is strongly anisotropic. This happens since the Rashba spin-orbit interaction is much weaker than the Dresselhaus spin-orbit interaction, $\gamma_R\ll\gamma_D$, together with the fact that the out-of-plane confinement is much stronger than the in-plane confinement so that for all relevant values of the magnetic field, $B_xf_\alpha,\,B_yf_\beta \ll g_c$ and $B_xf_\alpha',\,B_yf_\beta' \ll g_c'$. In the inset plot in Fig.~\ref{fig:D_32} we show the intra-valley coupling $\Delta_{21}$ as a function of the direction of the magnetic field at $B=1$ T for given positions of the interface steps. Since $\Delta_{21}$ solely originates from the coupling to the out-of-plane excited states, for a disordered quantum dot, we observe that in general it is much smaller than the inter-valley coupling, $\Delta_{32}$, and it vanishes in the absence of the magnetic field. As such, we neglect the inter-valley spin-valley coupling in the following analysis. In Appendix~\ref{app:dip_qlel} we explicitly show that neglecting $\Delta_{21}$ is well justified.

This enables us to obtain the following simplified relations for the modified qubit levels by only considering the coupling between states $|1\rangle$ and $|4\rangle$, and between $|2\rangle$ and $|3\rangle$,
\begin{align}\label{eq:qulv_1}
    |\tilde{1}\rangle&=\sqrt{\frac{1+a_+}{2}}e^{-i\mathrm{arg}[\Delta_{32}]}|1\rangle - \sqrt{\frac{1-a_+}{2}}|4\rangle,\\\label{eq:qulv_2}
    |\tilde{2}\rangle&=\sqrt{\frac{1-a_-}{2}}|2\rangle - \sqrt{\frac{1+a_-}{2}}e^{-i\mathrm{arg}[\Delta_{32}]}|3\rangle,\\\label{eq:qulv_3}
    |\tilde{3}\rangle&=\sqrt{\frac{1+a_-}{2}}|2\rangle + \sqrt{\frac{1-a_-}{2}}e^{-i\mathrm{arg}[\Delta_{32}]}|3\rangle,
\end{align}
in which  $a_{\pm}=\delta_{\pm}/\sqrt{\delta_{\pm}^2 +4|\Delta_{32}|^2}$, $\delta_{\pm}=E_z\pm E_{vs} $. To arrive to the above relations, we have neglected the coupling between states with the same spin direction (e.g. the coupling between $|1\rangle$ and $|3\rangle$). It is easy to show that the corrections to the qubit levels due to such same-spin couplings result in a sub-leading contribution to the spin relaxation. 

\begin{figure}[t!]
\begin{center}
\includegraphics[width=.25\textwidth]{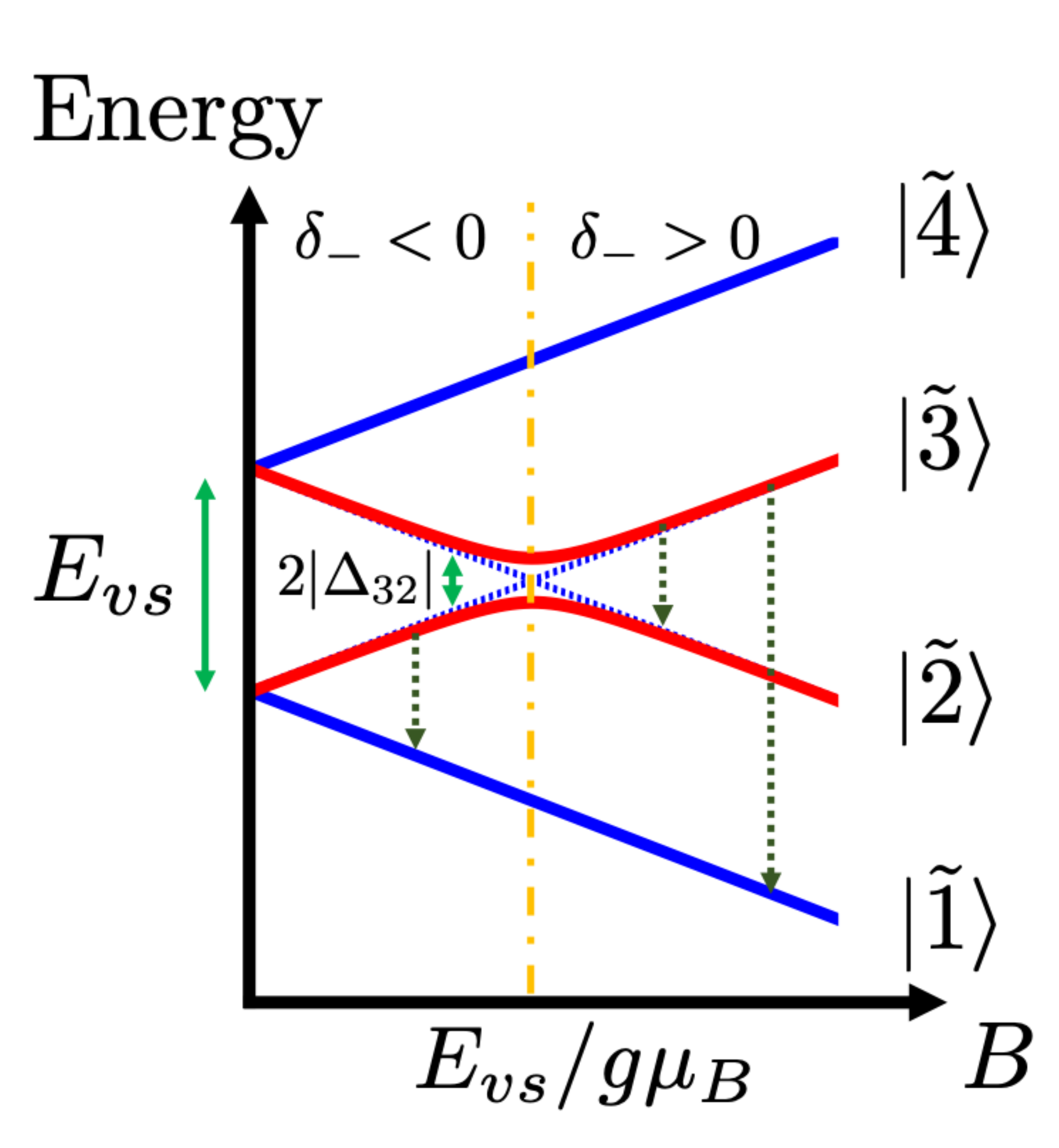}
\end{center}
\caption{Level diagram of a single-electron silicon spin qubit in the presence of SVM. The dot-dashed line highlights the magnetic field at which the hotspot occurs. The dotted arrows show the decay channels made possible by the spin-valley coupling $\Delta_{32}$. The decay rates are studied in 
Sec.~\ref{sec:qr}.}  \label{fig:lel_dig}
\end{figure}

For the low magnetic fields where $\delta_-<0$, the qubit logical excited state is $|\tilde{2}\rangle$ whereas for higher fields at which $\delta_->0$ the qubit logical excited state is $|\tilde{3}\rangle$, see Fig.~\ref{fig:lel_dig}. In both cases, the component of the qubit excited state including the state $|3\rangle$ enables the qubit relaxation. Likewise, the component including state $|4\rangle$ in the qubit ground state opens a decay channel. Note that these components are present only if the inter-valley coupling is finite ($\Delta_{32}\neq 0$), while the coefficient of state $|3\rangle$ in the qubit excited state reaches its maximal value at $B=E_{vs}/g\mu_B$. 

 This condition defines the spin-valley hotspot at which the qubit relaxation time is severely reduced. This is well studied in several experiments \cite{Yang13, Petit18, Borjans19, Zhang20,Hollmann20} while the excited states \esref{eq:qulv_2} and (\ref{eq:qulv_3}) are also already given by a number of previous works \cite{Yang13, Huang14,Petit18,Petit20,Huang20}. However, in sharp contrast with previous works where the spin-valley coupling $\Delta_{32}$ has been treated as a fitting parameter, the valley-dependent envelope function theory that we developed earlier in Sec. \ref{sec:VDEnv} enabled us to calculate $\Delta_{32}$ as a function of the interface roughness, and study its anisotropic properties. 
\subsection{Qubit-level modification due to SOM}\label{sec:SQL_SOM}
We now turn to study the corrections to the qubit levels due to the coupling to the orbital excited states induced by the spin-orbit interaction. The SOM becomes important at sufficiently high magnetic fields when the coupling between the levels $|1\rangle$ and $|3\rangle$ and the orbital excited states dominates over the inter-valley couplings. Moreover, at the leading order we can also neglect the corrections  to the orbital states due to interface steps. We then find the  modified qubit ground and excited states due to SOM,
\begin{align}\label{eq:ql_SOM_g}
    |\tilde{g}\rangle &\simeq |0,\downarrow\rangle + c_1|1_x,\uparrow\rangle + c_2|1_y,\uparrow\rangle,\\\label{eq:ql_SOM_e}
    |\tilde{e}\rangle &\simeq |0,\uparrow\rangle + c_3|1_x,\downarrow\rangle + c_4|1_y,\downarrow\rangle.
\end{align}
Here the unperturbed orbital states, i.e in the absence of spin-orbit interaction and assuming two valley states are degenerate, can be written as,
\begin{align}\label{eq:sol_0}
    |0\rangle &= e^{ik_0z}u_{+z}\psi_{x,0}\psi_{y,0}\psi_{z,0},\\\label{eq:sol_1x}
    |1_x\rangle &= e^{ik_0z}u_{+z}\psi_{x,1}\psi_{y,0}\psi_{z,0},\\\label{eq:sol_1y}
    |1_y\rangle &= e^{ik_0z}u_{+z}\psi_{x,0}\psi_{y,1}\psi_{z,0},
\end{align}
and the coefficients $c_1$ to $c_4$ are presented in Appendix~\ref{app:som}. We note that the coefficients $c_1$ and $c_2$ are proportional to $(\hbar\omega_x'+E_z)^{-1}$ and $(\hbar\omega_y'+E_z)^{-1}$ and, therefore, they decrease with increasing  magnetic field. However,  $c_{3}$ and $c_{4}$ are proportional to $(\hbar\omega_{x}' - E_z)^{-1}$ and $(\hbar\omega_{y}' - E_z)^{-1}$ so that the spin-down-component of the qubit excited state grows when increasing the magnetic field (for all practical values of magnetic field).

\section{Qubit relaxation}\label{sec:qr}
Having studied the modifications to the spin qubit levels due to SVM and SOM in the previous section, we now turn to study the qubit decay rate. We consider the relaxation due to the electron-phonon interaction as well as the Johnson noise due to a lossy transmission line and the $1/f$ charge noise.

\subsection{Relaxation induced by electron-phonon interaction}
As silicon has a nonpolar and centrosymmetric lattice, there is no piezoelectric interaction. However, the deformation
potential is common to all semiconductors, and it gives rise to an energy shift of the electronic states in the presence of lattice deformations. 
The electron-phonon deformation potential
interaction can be described by the following Hamiltonian,
\begin{align}
    H_\mathrm{e-ph}=H_\mathrm{e-ph}^l+H_\mathrm{e-ph}^t
\end{align}
where the contributions from the longitudinal and transverse phonons are,
\begin{align}
    H_\mathrm{e-ph}^l=i\sum_{q}&\sqrt{\frac{\hbar q}{2\rho_\mathrm{si}Vv_l}}\left(\Xi_d+\Xi_u\cos^2\theta_q\right)\nonumber\\
    &\times\left(b_{q,l}-b^\dagger_{-q,l}\right)e^{i\bq.\br},\\
    H_\mathrm{e-ph}^t=-i\Xi_u\sum_{q}&\sqrt{\frac{\hbar q}{2\rho_\mathrm{si}Vv_{t}}}\cos\theta_q\sin\theta_q\nonumber\\
    &\times\left(b_{q,t}+b^\dagger_{-q,t}\right)e^{i\bq.\br},
\end{align}
respectively.  Here, the phonon wavevector is $\bq=q(\cos\phi_q\sin\theta_q,\sin\phi_q\sin\theta_q,\cos\theta_q)$, $\rho_\mathrm{si} = 2330$ kg/m$^3$ is the mass density of silicon, $V$ is the volume, the deformation potential strengths are $\Xi_d=5$ eV and $\Xi_u=8.77$ eV, 
and the sound velocities for silicon amount to $v_l=9330$ m/s and $v_{t}=5420$ m/s. 

We now use  Fermi's golden rule to calculate the decay rate. We start by considering the spin-valley mixing and also use the electric dipole approximation $e^{i\textbf{q}.\textbf{r}}\simeq 1+ i\textbf{q}.\textbf{r}$ in the following analysis. We find for the qubit relaxation rate,
\begin{align}\label{eq:gamma_fi}
\Gamma_{|\tilde{f}\rangle\rightarrow |\tilde{i}\rangle}^{\mathrm{SVM, e-ph}} = \frac{E_z^5}{4\pi\rho_\mathrm{si}\hbar^6}(N(E_z)+1)\left(|x_{\tilde{f}\tilde{i}}|^2I_x + |z_{\tilde{f}\tilde{i}}|^2I_z\right),
\end{align}
where $N(E_z)=(e^{E_z/k_BT}-1)^{-1}$, and $r_{\tilde{f}\tilde{i}}=\langle \tilde{f}|\textit{r}|\tilde{i}\rangle$ ($r=x,z$) are the dipole moments between qubit states. Given the modified qubit states due to SVM discussed in Sec.~\ref{sec:SQL_SVM} and the inter-valley and intra-valley dipole moments we found in Sec.~\ref{subsec:dip} we can readily calculate the dipole moment between the qubit states as a function of the quantum dot parameters, see Appendix~\ref{app:dip_qlel} for details. We also defined,
\begin{align}
    I_x= &\frac{1}{v_l^7}\left[\frac{2}{3}\Xi_d^2 + \frac{4}{15}\Xi_d\Xi_u + \frac{2}{35}\Xi_u^2\right] + \frac{1}{v_t^7}\frac{8}{105}\Xi_u^2,\\
    I_z= &\frac{1}{v_l^7}\left[\frac{2}{3}\Xi_d^2 + \frac{4}{5}\Xi_d\Xi_u + \frac{2}{7}\Xi_u^2\right] + \frac{1}{v_t^7}\frac{4}{35}\Xi_u^2. 
\end{align}

As mentioned in the previous section, below the spin-valley hotspot where $\delta_{-}<0$, the qubit logical excited states is $|\tilde{f}\rangle=|\tilde{2}\rangle$ whereas above the spin-valley hotspot, the qubit logical excited states is $|\tilde{f}\rangle=|\tilde{3}\rangle$. In addition to the $|\tilde{3}\rangle \rightarrow |\tilde{1}\rangle$ relaxation in this case, there is an additional decay through $|\tilde{3}\rangle \rightarrow |\tilde{2}\rangle$, and we can again use \eref{eq:gamma_fi} to find its corresponding decay rate by replacing $E_z$ by $\delta_-$.

Let us now study the relaxation due to spin-orbit mixing for which the qubit levels are given by \esref{eq:ql_SOM_g} and (\ref{eq:ql_SOM_e}). We note that as the magnetic field becomes larger, the electric dipole approximation becomes less accurate due to the phonon-bottleneck effect \cite{Tahan14}. Therefore, we retain all multipoles and find for the relaxation rate,
\begin{align}\label{eq:gamma_SOM_eph}
    \Gamma_{|\tilde{e}\rangle\rightarrow |\tilde{g}\rangle}^\mathrm{SOM,e-ph} &=\frac{E_z^3}{8\pi^2\rho_\mathrm{si}\hbar^4}(N(E_z)+1)\nonumber\\
    &\times\left(\frac{\Xi_d^2I_0 + 2\Xi_d\Xi_uI_2+\Xi_u^2I_4}{v_l^5}+\frac{\Xi_u^2J}{v_t^5}\right)
\end{align}
in which we defined,
\begin{align}
    I_n&=|c_1^*+c_3|^2\int d\phi_q\int d\theta_q\sin\theta_q\cos^n\theta_{q}|\langle 1_x|e^{i\textbf{q}_l.\textbf{r}}|0\rangle|^2\nonumber\\
    &+|c_2^*+c_4|^2\int d\phi_q\int d\theta_q\sin\theta_q\cos^n\theta_{q}|\langle 1_y|e^{i\textbf{q}_l.\textbf{r}}|0\rangle|^2,\\
    J&=|c_1^*+c_3|^2\int d\phi_q\int d\theta_q\sin^3\theta_q\cos^2\theta_{q}|\langle 1_x|e^{i\textbf{q}_t.\textbf{r}}|0\rangle|^2\nonumber\\
    &+|c_2^*+c_4|^2\int d\phi_q\int d\theta_q\sin^3\theta_q\cos^2\theta_{q}|\langle 1_y|e^{i\textbf{q}_t.\textbf{r}}|0\rangle|^2,
\end{align}
where $\textbf{q}_{l(t)}=\frac{E_z}{\hbar v_{l(t)}}(\cos\phi_q\sin\theta_q,\sin\phi_q\sin\theta_q,\cos\theta_q).$ 
In the definitions of $I_n$ and $J$ given above, the term proportional to $|(c_1^* +c_3)(c_2^* +c_4)|$ is absent; this is due to the fact that this term vanishes when integrating over $\phi_q$. We also note that in order to obtain the formal definition of the $T_1$ time where both relaxation and excitation are possible, we substitute $N\rightarrow 2N$ in \esref{eq:gamma_fi} and (\ref{eq:gamma_SOM_eph}).

\subsection{Relaxation induced by Johnson noise and $1/f$ charge noise}
The Johnson noise is caused by the electromagnetic fluctuations in an electrical circuit. Such electromagnetic fluctuations, in turn, are generated by the thermal agitation of the charge carriers \cite{Weiss08}. In particular, the electron reservoir  used for loading and unloading the quantum dot is a relevant source for the Johnson noise \cite{Hollmann20, Petit18}. The electric noise spectra for such a lossy transmission line (LTR) is studied in \cite{Hollmann20} and it reads,
\begin{align}\label{eq:S_Ez}
    S^{\mathrm{J}}_E(\omega)= \frac{1}{\pi}\mathrm{D}_0\frac{1}{\sqrt{2}}\hbar\sqrt{\omega}\coth{\left(\frac{\hbar\omega}{2k_BT_\mathrm{el}}\right)}.
\end{align}
Here, we defined $\mathrm{D}_0= l_0^{-2}\sqrt{R/C}$ in which $l_0$ is the length scale between the source and drain,  $R$ and $C$ are the resistance and capacitance per unit length, and $T_\mathrm{el}$ is electron temperature in the reservoir. We can consider $\mathrm{D}_0$ as a fitting parameter whereas $T_\mathrm{el}$ is assumed to be known from experiment. 

Another possible electric noise is the $1/f$ charge noise that is generally known to originate, e.g.,  from the fluctuating two-level systems in the vicinity of the Si quantum well. The electric charge noise spectra can in general be written as,
\begin{align}
    S^{1/f}_E(\omega)=\frac{S_0}{\omega^\alpha},
\end{align}
in which $S_0$ determines the power spectral density at 1 Hz and the exponent $\alpha$ is device-dependent and it is typically reported to be between 0.5 and 2 \cite{Kranz20}.

Given the electric noise spectral function from the Johnson noise and $1/f$ charge noise, we can calculate the resulting qubit relaxation rate by using 
\begin{align}\label{eq:G_fi}
 1/T_1=\frac{4\pi e^2}{\hbar^2} S_E(\omega)\sum_{j}|\langle \tilde{f}|r_j|\tilde{i}\rangle|^2,
\end{align}
where $S_E$ can denote either the Johnson or $1/f$ noise power, or a combination of both, and where the form of the initial and final states, $|\tilde{i}\rangle$ and $|\tilde{f}\rangle$, depends on whether we consider spin-valley mixing or the spin-orbit mixing, as discussed in Sec.~\ref{sec:SQL}. Here $r=(x,y,z)$ and we recall that in our model the dipole moment of $y$ vanishes for qubit levels obtained due to SVM, whereas the dipole moment of $z$ vanishes for qubit levels obtained due to SOM.

\section{Discussion}\label{sec:dis}

In the previous sections, we developed a theory that, for a given interface roughness and electromagnetic field, predicts the valley splitting, the dipole matrix elements and the spin-valley couplings $\Delta_{32}$ and $\Delta_{21}$. As we showed in Sec.~\ref{sec:qr}, all of these quantities influence the qubit states and are therefore important in understanding the qubit relaxation time. 
In this section, we first show that our theory can faithfully reproduce and explain the experimental measurements presented in \ocite{Borjans19} with a minimal set of fitting parameters. We then proceed by investigating the behavior of the spin relaxation as a function of interface roughness, the direction of the magnetic field and the out-of-plane electric field.

\begin{figure}[t!]
\begin{center}
\includegraphics[width= 0.46\textwidth]{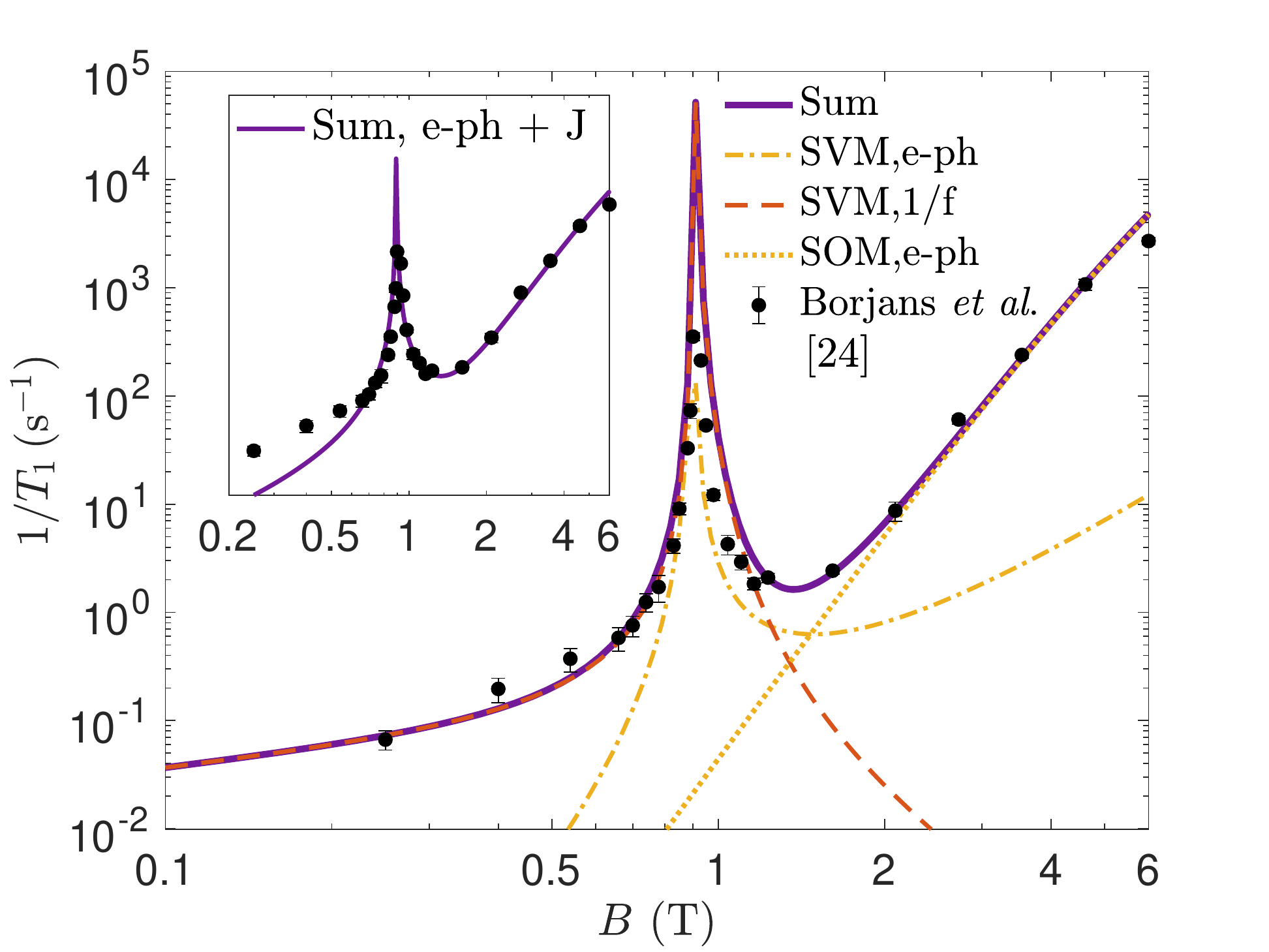}
\end{center}
\caption{Qubit relaxation rate $1/T_1$ as a function of magnetic field. Here the step positions are assumed to be at $x_{s\mathrm{L}}=-x_0$, $x_{s\mathrm{R}}=0.265x_0$.
The violet solid curve describes the total calculated $1/T_1$, whereas the other curves indicate the contributions due to electron-phonon interaction (orange dot-dahsed and yellow dotted) and the $1/f$ charge noise (red dashed). Experimental data points from \ocite{Borjans19} are shown as black circles. Following this reference, we set $\phi_B=\pi/4$, $\hbar\omega_x=3.9$ meV, and $T_\mathrm{el}=115$ mK. Inset: Comparison with Johnson instead of $1/f$ charge noise.}  \label{fig:T_1_the_vs_exp}
\end{figure}

In Fig.~\ref{fig:T_1_the_vs_exp} we show the theoretical prediction for the spin relaxation rate as a function of the magnetic field as well as the experimental data points from \ocite{Borjans19}. To obtain the theoretical result, we first searched for a set of locations for the interface steps $\{x_{s\mathrm{L}},x_{s\mathrm{R}}\}$ that gives rise to the same valley splitting energy as found from the experiment. Given the magnetic field where the spin-valley hotspot occurs, $B\simeq 0.91$ T, it turns out that the valley splitting amounts to $E_{vs}\simeq 105.2\,\mu$eV. Among a number of possibilities for the locations of the interface steps that give rise to this value for the valley splitting energy, we find that choosing $x_{s{\mathrm{L}}}=-x_0$ and $x_{s{\mathrm{R}}}=0.265x_0$ results in the best fit to the data.

At the next step, we consider the relaxation at high magnetic fields where the total relaxation rate is strongly dominated by the SOM and the electron-phonon interaction. By matching our model to the experimental data we find for the  parameter controlling the spin-orbit interaction strength  $A_0=5.42$. Afterwards, we consider the low B-field part of the data points where the qubit state modification due to SOM represents the dominant contribution to the spin relaxation. We find that considering the Johnson noise is not sufficient to explain the low B-field behavior of the spin relaxation, see the inset plot of Fig.~\ref{fig:T_1_the_vs_exp}. However, it is possible to fit the experimental data at the low B-fields by considering $1/$f charge noise. The fit shown in the main plot of Fig.~\ref{fig:T_1_the_vs_exp} is obtained by taking $S_0=10^{-3} (\mu \mathrm{V}/\mathrm{m})^2$ and $\alpha=1.5$. Assuming $l_0=100$ nm, this gives the amplitude of the voltage noise at $1$ Hz to be $10$ $\mu$eV$^2/$Hz which is within the range that has been reported for silicon quantum dots \cite{Kranz20}. We note that in \ocite{Huang20} another source for the Johnson noise is investigated in combination with considering the couplings between states $|1\rangle$ and $|4\rangle$, and similarly it is found that the low B-field behavior of the spin relaxation is determined by $1/$f charge noise.

\begin{figure}[t!]
\begin{center}
\includegraphics[width=0.5\textwidth]{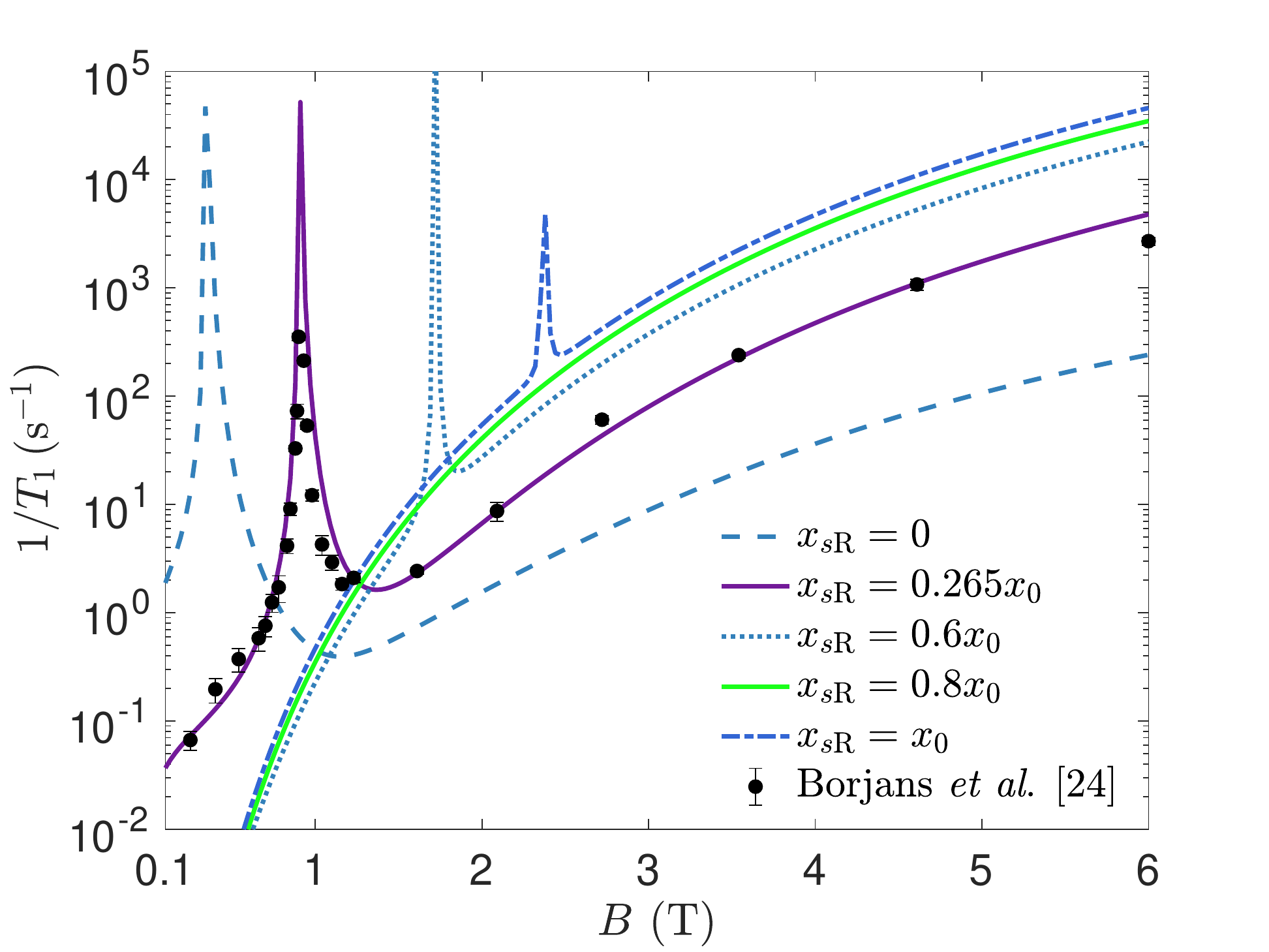}
\end{center}
\caption{The spin relaxation rate as a function of the magnetic field for various positions $x_{s\mathrm{R}}$ of one the interface steps. The experimental data is from \ocite{Borjans19}. Here we fixed the position of the other interface step $x_{s\mathrm{L}}=-x_0$, and the other parameters are the same as used in Fig.~\ref{fig:T_1_the_vs_exp}.  }  \label{fig:T1_vs_x_sR}
\end{figure} 
Having fixed these parameters, in Fig.~\ref{fig:T1_vs_x_sR} we study how the qubit relaxation time is changed by moving the position of the step at $x_{s\mathrm{R}}$. Let us first consider the low field region below the hotspot where the SVM and the 1$/f$ charge noise give the dominant contribution to the relaxation rate. We assume the magnetic field is sufficiently away from the hotspot so that we have $|\Delta_{32}|\ll|\delta_{-}|$. As such, we simplify \eref{eq:G_fi} and find at the leading order,
\begin{align}
\label{eq:J_SVM_21_simplified}
    \Gamma_{|\tilde{2}\rangle\rightarrow |1\rangle}^{\mathrm{SVM,1/f}}\propto S^{\mathrm{1/f}}_E(\omega)|\Delta_{32}|^2\left(\frac{1}{|\delta_-|}-\frac{1}{\delta_+}\right)|x_{10}|^2,
\end{align}
where we have neglected the $z_{10}$ dipole matrix element, justified from values shown in Fig.~\ref{fig:r_ij}. As shown there, by moving $x_{s\mathrm{R}}$ away from the quantum dot center, the $|x_{10}|$ inter-valley dipole matrix element decreases whereas the valley splitting increases. Changing $x_{s\mathrm{R}}$ also strongly influences the spin-valley coupling $\Delta_{32}$ as shown in Fig.~\ref{fig:D_32}. As a result, in Fig.~\ref{fig:T1_vs_x_sR} we observe that by increasing $x_s$, the decay rate at low magnetic fields below the hotspot decreases while the hotspot occurs at higher magnetic fields. Remarkably, at $x_s=0.8x_0$ we do not observe the spin-valley hotspot which is due to lack of spin-valley coupling, see Fig.~\ref{fig:D_32}. 

On the other hand, we find that by increasing $x_{s\mathrm{R}}$, the decay rate at  high magnetic fields above the hotspot increases. This is due to the behavior of the Dresselhaus term, \eref{eq:H_D},  together with the fact that $\gamma_R\ll \gamma_D$. As we mentioned earlier in Sec.~\ref{sec:SOI}, the coefficient of the Dresselhaus spin-orbit interaction changes sign by encountering a single-layer atomic step. As such, by moving the position of the atomic step away from the quantum dot center, the spatially averaged value $\langle\gamma_{D}\cos(4\pi z/a_0)\rangle$ grows. This, in turn, increases the coefficients $c_1$ to $c_4$ that quantify the correction to the qubit levels due to SOM in \esref{eq:ql_SOM_g} and (\ref{eq:ql_SOM_e}). See Appendix~\ref{app:som} for the exact mathematical expression of the coefficients $c_1$ to $c_4$.

We now turn to study how the qubit relaxation time depends on the out-of-plane electric field. Changing $F_z$ alters the out-of-plane envelope function at the interface, \eref{eq:psi_zAT0}. This directly modifies the corrections to the qubit levels due to the SVM and SOM. For the latter, the coefficients $c_1$ to $c_4$ in \esref{eq:ql_SOM_g} and (\ref{eq:ql_SOM_e}) at the leading order only involve the ground state of the out-of-plane motion. Using \eref{eq:psi_zAT0}, we realize that these coefficients scale linearly with the electric field. For the SVM, the scaling of the spin-valley coupling $\Delta_{32}$ with the electric field actually depends on the direction of the magnetic field. At $\phi_B=\pi/2$ in our model, the dominant contribution to $\Delta_{32}$ comes from the ground state of the out-of-plane motion, leading to a linear scaling. At $\phi_B=0$, however, one needs to rely on numerical analysis since a sizeable contribution of $\Delta_{32}$ involves excited states of the out-of-plane motion.

In addition, using  \esref{eq:psi_0z} and (\ref{eq:psi_zAT0}), one can see that at the first order in perturbation, the valley splitting from \eref{eq:Delta_1_dis} scales linearly with the electric field. This linear dependence is already observed in experiment \cite{Yang13,Ibberson18} and it was also previously predicted from theoretical analysis \cite{Hosseinkhani20,Ruskov18}. Based on the same reason, the in-plane dipole matrix elements also scale linearly with the electric field, see Fig.~\ref{fig:r_ij_vs_F_z_Ex}(a). In Fig.~\ref{fig:T_1_vs_Fz} we show the obtained qubit relaxation time for some fixed values of the electric field. As expected, the magnetic field at which the spin-valley hotspot accrues is reduced when decreasing the electric field. Interestingly, this feature enables us to turn the spin-valley hotspot into a ``coldspot" by properly reducing the electric field. In Fig.~\ref{fig:T_1_vs_Fz}, the vertical dashed lines highlight this possibility. We observe that the spin-valley hotspot obtained at $F_z=15\,(11)$ MV/m becomes a coldspot if the electric field is lowered to $F_z=8\,(4)$ MV/m. Note that at this coldspot, the relaxation rate is also first-order insensitive to the fluctuations of the magnetic field. We stress that while we can tune the electric field to increase the relaxation time, 
this electrical tunability also enables us to significantly shorten the relaxation time on demand, which allows for fast qubit reset and initialization, which is of crucial importance in performing quantum error correction \cite{Terhal15}. In the inset of Fig.~\ref{fig:T_1_vs_Fz}, we show the magnetic field at which spin-valley hotspot and coldspot occurs as a function of the electric field. One should bear in mind that these quantities depend of the interface roughness, and they would change by having some other positions for the interface steps.

\begin{figure}[b!]
\begin{center}
\includegraphics[width=.46\textwidth]{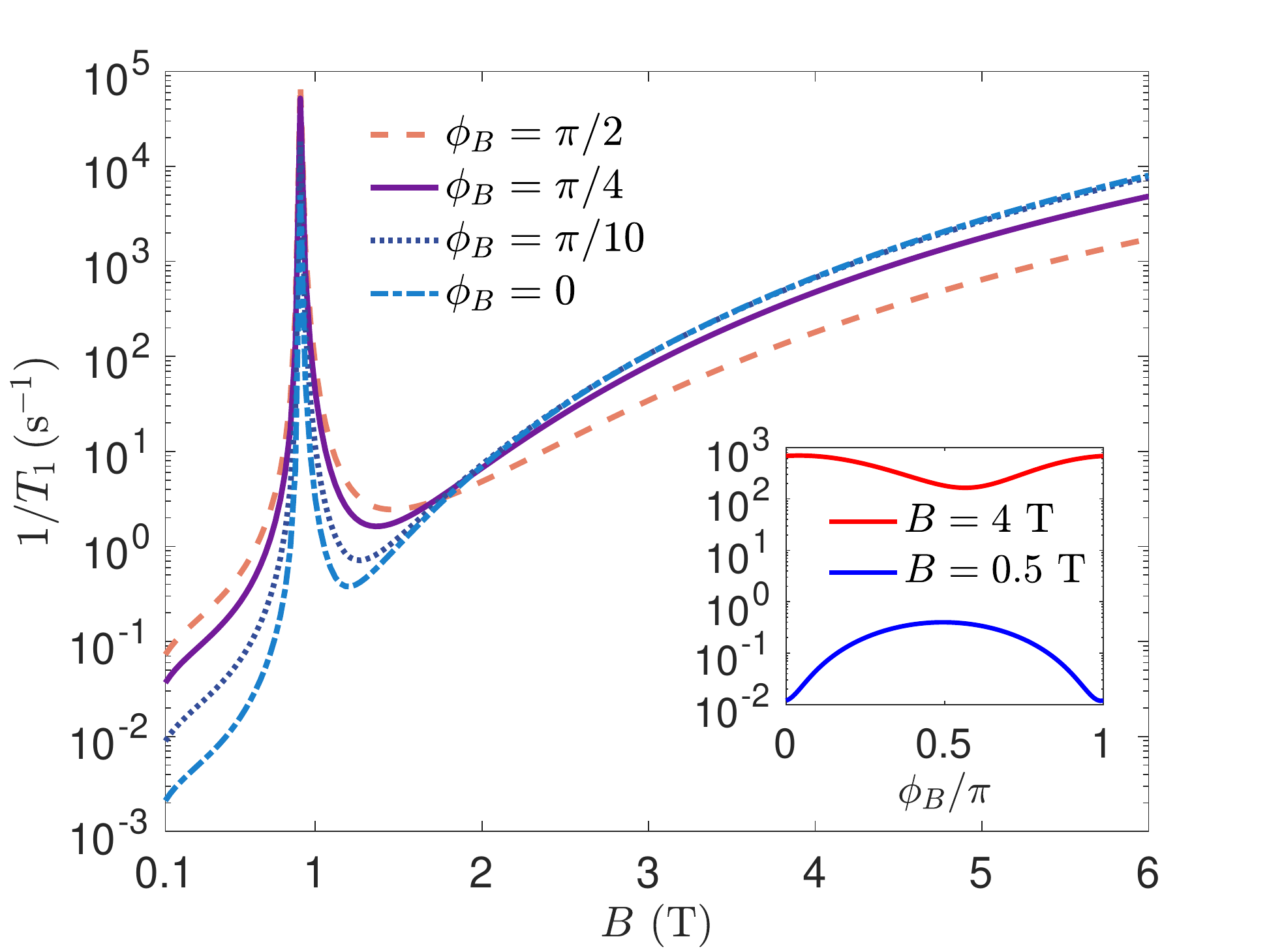}
\end{center}
\caption{The qubit relaxation rate as a function of magnetic field for different directions $\phi_B$ of the in-plane magnetic field. Inset: Qubit relaxation rate as a function of the direction of the magnetic field for fixed $B$. All the other parameters are the same as used in Fig.~\ref{fig:T_1_the_vs_exp}. }  \label{fig:T_1_vs_phi_B}
\end{figure}

Finally, we consider the anisotropic behavior of the spin relaxation. In Fig.~\ref{fig:T_1_vs_phi_B}, we use the same interface roughness as in Fig,~\ref{fig:T_1_the_vs_exp}, and observe a strong dependence of the spin relaxation rate on the direction of the magnetic field (described by the angle $\phi_B$ in the plane).
As discussed in Sec.~\ref{sec:SQL_SVM}, the inter-valley spin-valley coupling $\Delta_{32}$ given by \esref{eq:D_32} is strongly anisotropic for a quantum dot with disordered interface, see Fig.~\ref{fig:D_32}. Therefore, at low magnetic fields, where the SVM is the dominant decay channel, the qubit relaxation shows the same anisotropic response as $\Delta_{32}$. We recall that away from $\phi_B=\pi/4$, the intra-valley coupling  $\Delta_{21}$ becomes finite. However, its effect remains small for the full range of magnetic fields, see Fig.~\ref{fig:r_q_ij} in Appendix~\ref{app:dip_qlel}.

At higher  magnetic fields we observe a different anisotropic behavior in Fig.~\ref{fig:T_1_vs_phi_B}. This is because at high fields the SOM becomes the dominant decay channel, see Fig.~\ref{fig:T_1_the_vs_exp}, and because, within our model where the steps are assumed parallel to the $\hat{y}$-axis, we have $\langle 1_x|\{p_x,\mathcal{S}_\mathrm{int}\}|0\rangle\leq \langle 1_y|\{p_y,\mathcal{S}_\mathrm{int}\}|0\rangle$, where the equality (for a circular dot) is reached only when the steps are far away from the dot center (i.e. when the quantum dot is ideally flat), see \esref{eq:px_1x0} and (\ref{eq:py_1y0}). This inequality in the presence of interface steps gives rise to $c_3(\phi_B=0)\ll c_4(\phi_B=0)$ and $c_4(\phi_B=\pi/2)\ll c_3(\phi_B=\pi/4)< c_4(\phi_B=0)$ which, in turn, leads to the anisotropic behavior of the spin relaxation at high magnetic fields above the hotspot. In the inset of Fig.~\ref{fig:T_1_vs_phi_B} we show the spin relaxation rate as a function of the direction of the magnetic field for two fixed values of $B$ below and above the hotspot. We find a change of nearly two orders of magnitude in the $T_1$ time for $B=0.5$ T. A similarly large effect has recently been reported in experiment, see \ocite{Zhang20}.

\section{Conclusions and outlook}\label{sec:conc}

Silicon spin qubits are among the most promising platforms for scalable quantum computation. However, the presence of two low-lying valley states in silicon quantum dots can potentially be harmful for Si spin qubits. It has been known that the presence of interface steps can render the valley structure and with it the relaxation ($T_1$) time characteristics of Si spin qubits sample-dependent. This, in turn, can pose a challenge to the scalability of silicon-based platforms. On the other hand, to the best of our knowledge, so far there has not been a general theory to predict the behavior of spin relaxation as a function of the interface roughness. In this paper, we achieved this by first developing a valley-dependent envelope function theory in Sec.~\ref{sec:VDEnv} that can predict the valley splitting, the dipole matrix elements and the spin-valley coupling for a given interface roughness and the electromagnetic fields.

Our approach enables us to substantially reduce the number of free parameters in the theory of qubit relaxation. For the sake of simplicity, throughout this work we assumed that the interface roughness is stair-like, and the steps are formed parallel to the $\hat{y}$-axis, as schematically depicted in Fig.~\ref{fig:dis_dot}. However, it is easy to generalize our perturbative treatment of the interface roughness to any arbitrary configuration of the interface steps. In Sec.~\ref{sec:SOI}, we formulate a general form for the interface-induced spin-orbit interaction for a 3D electron. Based on this description, we find the corrections to the spin qubit levels due to spin-valley and spin-orbit coupling in Sec.~\ref{sec:SQL}. Remarkably, we discovered that under certain conditions for a disordered quantum dot, the spin-valley coupling can vanish. This can have a major effect of the qubit lifetime as it completely blocks the valley-induced decay. Moreover, our analysis also allows us to investigate the anisotropic behavior of the spin-valley coupling, see Fig.~\ref{fig:D_32}.

In Sec.~\ref{sec:qr} we consider the electron-phonon interaction and the Johnson and $1/f$ charge noise and discussed how these mechanisms give rise to qubit relaxation. Finally, in Sec.~\ref{sec:dis} we present our final results for the qubit relaxation time $T_1$. In Fig.~\ref{fig:T_1_the_vs_exp} we show that our theory can well reproduce  experimental data for the qubit relaxation \ocite{Borjans19} with only a minimal set of free parameters. We found that in order to fit the data at low B-fields, it is necessary to include the effects of the $1/f$ charge noise.  We also investigated how the qubit relaxation rate changes when one step that is close to the dot center is moved away, see Fig.~\ref{fig:T1_vs_x_sR}. Here we show that the spin-valley hotspot disappears when the spin-valley coupling vanishes.

We further studied how the qubit relaxation depends on the direction of the magnetic field. We find that the presence of the interface steps can give rise to a strong anisotropic behavior. Within our model, we find that the $T_1$ time can either increase or decrees by changing the magnetic field, depending on whether SVM or SOM represents the dominant decay channel, see Fig.~\ref{fig:T_1_vs_phi_B}. We finally studied how the out-of-plane electric field $F_z$ generated by the gate voltages influences the qubit relaxation time, see Fig.~\ref{fig:T_1_vs_Fz}. We find that the relaxation rate can vary by several orders of magnitudes when $F_z$ is changed. 

At a fixed magnetic field, there is an optimal electric field that sets the qubit on a coldspot where the $T_1$ time reaches a local maximum and becomes first-order insensitive to the fluctuations of $B$. Importantly, the electric field can also be tuned to set the qubit at the spin-valley hotspot. This, in turn, enables an on-demand qubit reset which is necessary, and of great importance, for scalable quantum computation. While the presence of the valley degree of freedom in silicon heterostructures has so far been commonly viewed as a problematic feature, we therefore demonstrate that upon proper control over the out-of-plane electric field, the spin-valley coupling can in fact be an advantage for silicon-based platforms. 

In conclusion, we point out that while we employed the valley-dependent envelope function theory for analyzing a single electron Si/SiGe quantum dot, our theory can also be applied for a disordered Si/SiO$_2$ quantum dot by using the offset potential of the SiO$_2$ ($U_0^{\mathrm{SiO}_2}$=3 meV), and with the knowledge of the  periodic parts of the Bloch function in SiO$_2$. The latter is studied in \ocite{Koiller11}; however, based on the same reference, further studies may be required for a more accurate understanding. 
\acknowledgments
We gratefully acknowledge useful discussions with J. R. Petta, F. Ginzel and M. Russ. This work has been supported by ARO grant number W911NF-15-1-0149.

\appendix
\section{The valley-dependent envelope function: perturbative method}\label{app:PC}
Here we present how to solve \eref{eq:VDenv} in order to arrive to the valley-dependent envelope function \eref{eq:Psi_t_p} in the presence of valley-coupling, interface steps and an in-plane magnetic field. To begin with, we assume for the moment that the valley-dependent envelope functions $\Psi_{xyz}^{\pm z}$ are known.  We then start from \eref{eq:VDenv} and first multiply it by $(\Psi_{xyz}^{+z})^*e^{-ik_0z}$ followed by an integration over the spatial coordinates. This, at the leading order, leads us to
\begin{align}\label{eq:p1}
    a_{+z}\left[E_{+z} + \Delta_{+z} \right] + a_{-z}\Delta_1 = a_{+z}E,
\end{align}
where we defined,
\begin{align}
    &E_{+z}=\int (\Psi_{xyz}^{+z})^*H_c\Psi_{xyz}^{+z}d^3r,\\\label{eq:D+z}
    &\Delta_{+z} = \int (\Psi_{xyz}^{+z})^*V_v(r)\Psi_{xyz}^{+z}d^3r,\\\label{eq:D1}
    & \Delta_1 = \int e^{-2ik_0z}(\Psi_{xyz}^{+z})^*V_v(r)\Psi_{xyz}^{+z}d^3r.
\end{align}
In a similar way, we multiply \eref{eq:VDenv} by $(\Psi_{xyz}^{-z})^*e^{ik_0z}$ and perform an integration to find,
\begin{align}\label{eq:p2}
    a_{+z}\Delta_1^* +a_{-z} \left[E_{-z} + \Delta_{-z} \right]  = a_{-z}E,
\end{align}
where $E_{-z}$ and $\Delta_{-z}$ are defined similar to \esref{eq:D+z} and (\ref{eq:D1}) by using $\Psi_{xyz}^{-z}$. Later on, it turns out that $E_{-z}=E_{+z}$ and $\Delta_{-z}=\Delta_{+z}$ 

We now use these relations in order to recast \esref{eq:p1} and (\ref{eq:p2}) into a matrix equation,
\begin{align}\label{eq:M_E}
    \begin{bmatrix} E_{+z} + \Delta_{+z} & \Delta_1\\ \Delta_1^* & E_{+z} + \Delta_{+z} \end{bmatrix}\begin{bmatrix} a_{+z}\\a_{-z}\end{bmatrix}=E \begin{bmatrix} a_{+z}\\a_{-z}\end{bmatrix}.
\end{align}
Solving \eqref{eq:M_E} yields for the  energies of the valley-orbital ground ($q=0$) and excited ($q=1$) states,
\begin{align}\label{eq:E_q}
    E^{(q=0,1)}=E_{+z}+\Delta_{+z} \mp |\Delta_1|,
\end{align}
with the eigenvectors given by,
\begin{align}
    a_{\pm z}^{(q)}&=(\pm1)^{1-q}\frac{1}{\sqrt{2}}e^{\pm\frac{i}{2}\mathrm{arg}[\Delta_1]}.
    \label{eq:a_q}
\end{align}

We note here that all of the quantities on the right hand side of the equation (\ref{eq:E_q})  (i.e. $E_{+z},\:\Delta_{+z}$ and $\Delta_1$) in general turn out to be dependent on the valley-orbital index $q$. As we will see in the following, this dependence is due to the fact that the envelope function depends on the valley-orbital index so that in general $\Psi_{xyz}^{+z,(q=0)}\neq \Psi_{xyz}^{+z,(q=1)}$. The valley-splitting, i.e. the energy gap between the two low-lying valley-orbital states, then becomes
\begin{align}
\label{eq:Evs}
    E_{vs}=E^{(q=1)}-E^{(q=0)}.
\end{align}

In order to model the valley coupling parameter $V_v$, let us consider a quantum dot with an ideally flat interface at zero magnetic field. As we will see later, we can argue that due to the strong out-of-plane confinement caused by the electric field, the valley-dependent correction to the envelope function is negligible, enabling us to write $\Psi_{xyz}^{\pm z,(q=0)}\simeq \Psi_{xyz}^{\pm z,(q=1)} = \psi_{xyz,0}$.
In this case, as also explained in detail in \ocite{Friesen10}, the valley coupling becomes intra-orbital and the valley splitting, \eref{eq:Evs}, is simplified to,
\begin{align}
\label{eq:Evs_perf}
    E_{vs}^\mathrm{ideal}=2|\Delta_1^\mathrm{ideal}|.
\end{align}

Given the above equation, we set $\Delta_1^\mathrm{ideal}$ equal to the valley-orbit coupling for an ideal quantum dot, $\Delta_{vo}^\mathrm{ideal}$. The latter quantity is explained and discussed in detail in \ocite{Hosseinkhani20}. It has been shown in that work that
\begin{align}
\label{eq:D_vo_perf}
\Delta_{vo}^\mathrm{ideal}=& \langle\psi_{xyz,0}u_{+z}(r)e^{ik_0z}|U_0\theta(z)|\psi_{xyz,0}u_{-z}(r)e^{-ik_0z}\rangle\nonumber\\
=&-i\mathcal{C}_0\frac{eF_z}{2k_0}\left(1-\left[1-\frac{1}{2\tu_0}+i\frac{k_0z_0}{\sqrt{\tu_0}}\right]^{-1}\right).
\end{align}
We now use \esref{eq:psi_0z} and (\ref{eq:E_0z}) and take $V_v^\mathrm{ideal}(r)=V_v\delta(r)$ to find $\Delta_1^\mathrm{ideal} \simeq V_v/(z_0\tu_0)$. By setting this equal to \eref{eq:D_vo_perf}, we find a relation for the valley-coupling parameter,
\begin{align}\label{eq:Vv}
    V_v=z_0\tu_0\Delta_{vo}^\mathrm{ideal}.
\end{align}
We note here that, as explained in \ocite{Hosseinkhani20}, the quantity $\mathcal{C}_0$ in \eref{eq:D_vo_perf} has a microscopic nature and originates from the lattice-periodic parts of the Bloch function, $u_{\pm z}(r)$ in \eref{eq:vowf}. Based on the atomistic calculations performed in \ocite{Koiller11} for silicon, it is reported in \ocite{Hosseinkhani20} that $\mathcal{C}_0=\sum_{\bg}C_{+}^*(\bg)C_{-}(\bg)\simeq -0.2607$ where the reciprocal vector $\bg$ is introduced in \eref{eq:U_r_eq}. 

Having found the appropriate relation for the valley-coupling parameter \eref{eq:Vv}, we now consider the presence of the interface steps and magnetic field and study how the valley-dependent envelope function can be obtained. The out-of-plane potential in \eref{eq:H_xyz_B0} is modified by the presence of the interface steps and it becomes,
\begin{align}
\label{eq:Udis}
U_{\mathrm{dis}}(x,z)= U(z)+U_{\mathrm{steps}}(x,z),
\end{align}
where $U(z)$ is given by \eref{eq:Uz} and,
\begin{align}
\label{eq:U_pert}
U_{\mathrm{steps}}(x,z)&=U_0\theta(-z)\theta\left(z+\frac{a_0}{4}\right)\theta(x_{s\mathrm{L}}-x) \nonumber\\
  &-U_0\theta(z)\theta\left(z-\frac{a_0}{4}\right)\theta(x-x_{s\mathrm{R}}) .
\end{align}
Furthermore, we can write for the valley-coupling parameter $V_v^\mathrm{dis}(r)=V_v \mathcal{S}_\mathrm{int}(x,z)$ in which the interface function $\mathcal{S}_\mathrm{int}(x,z)$ vanishes everywhere except at the interface,
\begin{align}\label{eq:Sint}
    \mathcal{S}_\mathrm{int}(x,z)&=\delta\left(z+\frac{a_0}{4}\right)\theta(x_{s\mathrm{L}}-x)\nonumber\\
    &+\delta(z)\theta(x-x_{s\mathrm{L}})\theta(x_{s\mathrm{R}}-x)\nonumber \\
&+\delta\left(z-\frac{a_0}{4}\right)\theta(x-x_{s\mathrm{R}}).
\end{align}

In order to consider the presence of a homogeneous in-plane magnetic field, $\textbf{B}_{||}=(B_{x},B_{y},0)=B(\cos\phi_B,\sin\phi_B,0)$, similar to \ocite{Hosseinkhani20}, we  use the gauge where
$    \textbf{A}=[0,0,yB_x-xB_y]$.
The confinement Hamiltonian \eref{eq:H_xyz_B0} is accordingly modified by replacing $p_i\rightarrow p_i-eA_i$. We then find that in the presence of the interface steps and the in-plane magnetic field the confinement Hamiltonian becomes,
\begin{align} 
\label{eq:Htbp}
H_c= H_0' + H_{||} + U_{\mathrm{steps}}(x,z).
\end{align}
Here $H_0'$ is the separable and exactly solvable Hamiltonian of the same form as $H_0$ in \eqref{eq:H_xyz_B0} with $\omega_x$ and $\omega_y$ replaced by 
magnetic-field dependent confinement frequencies $\omega_x'$ and $\omega_y'$,
while the couplings induced by the in-plane magnetic field become,
\begin{align}
\label{eq:H_||}
H_{||}&=-B_x\frac{e}{m_l}yp_z+B_y\frac{e}{m_l}xp_z -B_xB_y\frac{e^2}{m_l}xy.
\end{align}
Let us first define the cyclotron frequency and magnetic length induced by  $B_{x(y)}$ by,
\begin{align}\label{eq:lB}
   &\Omega_{x(y)}=\frac{eB_{x(y)}}{\sqrt{m_tm_l}},
   &l_{x(y)}=\sqrt{\frac{\hbar}{eB_{x(y)}}}.
\end{align}
We can then write,
\begin{align}\label{eq:om}
&\omega_{x}'=\omega_{x}\left(1+\frac{\Omega_{y}^2}{\omega_{x}^2}\right)^{1/2},
&\omega_{y}'=\omega_{y}\left(1+\frac{\Omega_{x}^2}{\omega_{y}^2}\right)^{1/2}.
\end{align} 

In order to find the valley-dependent envelope function, we begin by rewriting \eref{eq:VDenv} in the presence of interface steps and the magnetic field as,
\begin{align}
\label{eq:VDenv_dis}
\sum_{j=\pm z}a_j^{(q)}e^{ik_jz}\left\{H_0'+H_{\mathrm{p}}-E^{(q)}\right\}\Psi_{xyz}^{j,(q)}=0,
\end{align}
where we defined,
\begin{align}\label{eq:H_p}
    H_{\mathrm{p}}= H_{||} +  V_v \mathcal{S}_\mathrm{int}(x,z) + U_{\mathrm{steps}}(x,z),
\end{align}
which can be considered  a perturbation due to the presence of the in-plane magnetic field, the valley-coupling and the interface steps. In the absence of $H_{\mathrm{p}}$, as explained before, the contribution from the two valleys become (nearly) decoupled. This enables us to simplify \eref{eq:VDenv_dis} leading to $H_0'\psi_{xyz}'=\e'\psi_{xyz}'$. This is the generalization of \eref{eq:schH0} in which the confinement Hamiltonian, and therefore the eigenenergies and eigenstates, are modified by the magnetic field. For simplicity, from now on, we drop the prime  $(')$ of the energy and eigenstate and bear in mind that $\psi_{x,m}$ and $\psi_{y,p}$ and their corresponding energies depend on the magnetic field, according to \esref{eq:om}. 

We now argue that since the eigenstates of $H_0'$ form a complete basis, in the presence of the perturbation $H_{\mathrm{p}}$, the solution for the ground $(q=0)$ and first excited $(q=1)$ valley-orbital envelope function of \eref{eq:VDenv_dis} can be described by the following general expansion,
\begin{align}
\label{eq:psi_fp}
    \Psi_{xyz}^{\pm z,(q)}= \psi_{xyz,0} + \sum_{m,p,n} c_{m,p,n}^{\pm z,(q)}\psi_{x,m}\psi_{y,p}\psi_{z,n},
\end{align}
where $\psi_{xyz,0}=\psi_{x,0}\psi_{y,0}\psi_{z,0}$ is the orbital ground state and thus $\{m,p,n\}=\{0,0,0\}$ is excluded from the summation. A similar disorder expansion is performed in \ocite{Gamble13} in combination with a tight binding method. Given \eref{eq:psi_fp}, the problem now simplifies to finding the coefficients $c_{m,p,n}^{\pm z,(q)}$. In the following, we aim to obtain these coefficients for the ground and first excited valley-orbital states up to the first order in the perturbation. As such, we write for the perturbed eigenenergy,
\begin{align}
\label{eq:e_fp}
    E^{(q)}=\e_0 + \delta E^{(q)}.
\end{align}

We now first substitute \esref{eq:psi_fp} and (\ref{eq:e_fp}) into \eref{eq:VDenv_dis}. Afterwards, we multiply the resulting relation by $e^{-ik_0z}\psi_{x,m'}\psi_{y,p'}\psi_{z,n'}\: ((m',p',n')\neq (0,0,0))$ followed by an integration over the spatial coordinate, $\mathbf{r}=(x,y,z)$. This enables us to find at the first order (and after changing $(m',p',n')\rightarrow (m,p,n)$),
\begin{align}
\label{eq:Eq_c_mn}
&a_{+z}^{(q)}\left\{(\e_{m,p,n}-\e_0)c_{m,p,n}^{+z,(q)}+ \mathcal{P}_{m,p,n}(H_\mathrm{p}) \right\}\nonumber\\
&+a_{-z}^{(q)}\Big\{\mathcal{F}_{m,p,n}(H_\mathrm{p})\Big\}=0.
\end{align}
where we defined the tensor elements,
\begin{align}\label{eq:PH_p}
    \mathcal{P}_{m,p,n}(H_\mathrm{p})&=\int \psi_{x,m}\psi_{y,p}\psi_{z,n}H_\mathrm{p}\psi_{xyz,0}d^3r,\\\label{eq:FH_p}
    \mathcal{F}_{m,p,n}(H_\mathrm{p})&=\int e^{-2ik_0z}\psi_{x,m}\psi_{y,p}\psi_{z,n}H_\mathrm{p}\psi_{xyz,0}d^3r.
\end{align}
Here the unperturbed eigenenergy reads,
\begin{align}
    \e_{m,p,n}=\left(\frac{1}{2}+m\right)\hbar\omega_x'+\left(\frac{1}{2}+p\right)\hbar\omega_y'+\e_{z,n},
\end{align}
while $\e_0=\e_{m,p,n}$ with $(m,p,n)=(0,0,0)$.
By using \eref{eq:a_q}  we then arrive to,
\begin{align}
\label{eq:c_mn_p}
c_{m,p,n}^{+z,(q)}&=\frac{(-1)^q e^{-i\mathrm{arg}[\Delta_1]}\mathcal{F}_{m,p,n}-\mathcal{P}_{m,p,n}}{\e_{m,p,n}-\e_0},
\end{align}

  In a similar way, after substituting \esref{eq:psi_fp} and (\ref{eq:e_fp}) into \eref{eq:VDenv_dis}, we multiply the resulting relation by $e^{+ik_0z}\psi_{x,m'}\psi_{y,p'}\psi_{z,n'}$ and perform the same calculations as presented here. We then find,
\begin{align}\label{eq:c_mn_m}
 c_{m,p,n}^{-z,(q)}&=\frac{(-1)^q e^{i\mathrm{arg}[\Delta_1]}\mathcal{F}_{m,p,n}^*-\mathcal{P}_{m,p,n}}{\e_{m,p,n}-\e_0}.
\end{align}
Note that at the first order in the perturbation, $\Delta_1$ is given by \eref{eq:Delta_1_dis}. Given \eref{eq:FH_p}, we find $\mathcal{F}_{m,p,n}(H_{||})\simeq 0$. This is due to the fast oscillations caused by the factor $e^{-2ik_0z}$ inside the integrand in \eref{eq:FH_p}. We thus realize from  \esref{eq:c_mn_p} and (\ref{eq:c_mn_m}) that the corrections to the envelope function due to the presence of the magnetic field are identical for both valley-orbital states, $q=0$ and $q=1$, so that we can drop the q-index of the corrections in this case. On the other hand, for the other two contributions that are both influenced by the presence and configuration of the interface steps, it is easy to confirm that in general $\mathcal{F}_{m,p,n}(V_v \mathcal{S}_\mathrm{int}(x,z) + U_{\mathrm{steps}}(x,z))\neq 0$. 

Indeed, (only) the corrections due to these interface-step-related terms depend on the valley-orbital index, $q$. Moreover, for such corrections to the envelope function, since in our model the steps are assumed to be parallel to the $\hat{y}$-axis, we find from \esref{eq:PH_p} and (\ref{eq:FH_p}) that $c_{m,p,n}^{\pm z, (q)}$ can be nonzero only by setting $p=0$. We then find the valley-dependent envelope function in the presence of the perturbations given by \eref{eq:H_p} (for $q=0$ and $q=1$ states) has the general form given by \eref{eq:Psi_t_p} in Sec.~\ref{sec:VDEnv}. 
\begin{table}[t!]
\begin{center}
\caption{\label{tab:par_alpha} The out-of-plane eigenenergies $\e_{z,n}$ as well as the coefficients $\alpha_n$ (in units of inverse Tesla). Here we consider a circular dot at $B=0$ (so that $\alpha_n=\beta_n$), and we set for the thickness of Si quantum well $d_t=10$ nm and for the thickness of upper SiGe barrier $d_b=46$ nm. The other used quantum dot parameters are given in the caption of Fig.~\ref{fig:DP_PD}. We also find from \eref{eq:eta} $\eta=-8.89\times 10^{-4}$ T$^{-2}$. }
\begin{tabular}{c c  c} \hline\hline
$n$ & $\e_{n,z}$ (meV) & $\alpha_n\: (10^{-3}\mathrm{T}^{-1})$  \\ \hline 
0 & 40.92  & 0     \\   
1 & 77.29  & -7.28 \\
2 & 106.84 & 2.25  \\ 
3 & 133.30 & -1.09 \\ \hline\hline
\end{tabular}
\end{center}
\end{table}

\begin{table*}[t!]
\begin{center}
\caption{\label{tab:par_cmn} The perturbative coefficients $c_{m,n}^{ +z,(q)}$ used in \eref{eq:psi_st}. The location of interface steps and other parameters are the same as given by caption of Fig.~\ref{fig:DP_PD}.}
\begin{tabular}{ c  c  c  c  c } \hline\hline
$m$ & $c_{m,0}^{+z,(q=0)}$                 & $c_{m,1}^{+z,(q=0)}$               & $c_{m,2}^{+z,(q=0)}$              &  $c_{m,3}^{+z,(q=0)}$             \\ \hline
$0$ & N/A                 & $-$0.0235 + 0.0138i &   0.0131 $-$0.0075i & $-$0.0091 + 0.0050i\\ 
$1$ & 0.2795 $-$0.0638i    & $-$0.0277 + 0.0065i &  0.0160 $-$0.0038i & $-$0.0111 + 0.0027i\\ 
$2$ & 0.0309 $-$0.0204i    & $-$0.0056 + 0.0035i &  0.0034 $-$0.0020i & $-$0.0024 + 0.0013i\\
$3$ & $-$0.0304 + 0.0123i   & 0.0076 $-$0.0030i  & $-$0.0047 + 0.0018i &  0.0034 $-$0.0013i\\ 
$4$ & $-$0.0140 + 0.0059i   & 0.0043 $-$0.0018i  & $-$0.0028 + 0.0011i &  0.0020 $-$0.0008i\\ \hline 
$m$ & $c_{m,0}^{+z,(q=1)}$ & $c_{m,1}^{+z,(q=1)}$  & $c_{m,2}^{+z,(q=1)}$ &  $c_{m,3}^{+z,(q=1)}$             \\ \hline
$0$ & N/A                 & $-0.0008 - 0.0041$i   & $0.0006 + 0.0022$i  & $-0.0005 - 0.0015$i\\ 
$1$ & $0.2205 + 0.0831$i & $-0.0210 - 0.0081$i   & $0.0116 + 0.0046$i  & $-0.0078 - 0.0032$i\\ 
$2$ & $-0.0827 + 0.0188$i &  $0.0142 - 0.0033$i   &$-0.0081 + 0.0019$i  & $0.0054 - 0.0013$i\\
$3$ & $0.0190 - 0.0129$i  & $-0.0045 + 0.0031$i   & $0.0026 - 0.0019$i  & $-0.0018 + 0.0013$i\\ 
$4$ & $0.0104 - 0.0061$i  &$-0.0030 + 0.0018$i    &$0.0018 - 0.0011$i   & $-0.0012 + 0.0008$i\\ \hline\hline
\end{tabular}
\end{center}
\end{table*}
Using $H_{||}$ and \esref{eq:PH_p} and (\ref{eq:c_mn_p}), we arrive to $\psi_{||}$ in \eref{eq:psi_par} in which the perturbative coefficients become,
\begin{align}
\label{eq:alpha_n}
\alpha_n&=-\frac{1}{2}\hbar\frac{e}{m_l}\frac{y_0'}{z_0}\frac{\langle\psi_{z,0}|\partial/\partial \tilde{z}|\psi_{z,n}\rangle}{\e_{z,0}-\e_{z,n}-\hbar\w_y'},\\ \label{eq:beta_n}
\beta_n&=-\frac{1}{2}\hbar\frac{e}{m_l}\frac{x_0'}{z_0}\frac{\langle\psi_{z,0}|\partial/\partial \tilde{z}|\psi_{z,n}\rangle}{\e_{z,0}-\e_{z,n}-\hbar\w_x'},\\\label{eq:eta}
\eta&=-\frac{1}{4}\frac{e^2}{m_l}x_0'y_0'\frac{1}{\hbar\w_x'+\hbar\w_y'},
\end{align}
where $\tz = z/z_0$. \esref{eq:alpha_n} and (\ref{eq:beta_n}) show that $\alpha_n=\beta_n$ for a circular dot when either $B=0$ or $\phi_B=\pi/4$ (and $B$ is nonzero). In other cases (and for quantum dots with realistic parameters), these coefficients are different but remain close to each other since the confinement along $\hat{z}$ in quantum dots is always stronger than the in-plane confinements. In Table~\ref{tab:par_alpha} and its caption, we show an example for the coefficients used in writing $\psi_{||}$.

We now move to consider the perturbative coefficients $c_{m,n}^{ +z,(q)}$ in \eref{eq:psi_st}. We can numerically find them by first calculating $\mathcal{P}_{m,0,n}(H_\mathrm{st})$ and $\mathcal{F}_{m,0,n}(H_\mathrm{st})$ from \esref{eq:PH_p} and (\ref{eq:FH_p}), followed by using \eref{eq:c_mn_p} and setting $p=0$. We remind that $c_{m,n}^{ -z,(q)}=[c_{m,n}^{+z,(q)}]^*$ as can be seem from \esref{eq:c_mn_p} and (\ref{eq:c_mn_m}). Table~\ref{tab:par_cmn} contains an example for the values of $c_{m,n}^{ +z,(q)}$ for the ground and excited valley-orbital states.

\section{Dipole matrix elements and their properties}\label{app:dipole}
Here we present some details of the calculations leading to the inter-valley and intra-valley dipole matrix elements presented in Sec.~\ref{subsec:dip}, and also study some properties of the valley-dependent dipole moments. Using \esref{eq:pmZ} and the plane wave expansion of the Bloch periodic part of the wave function \eref{eq:U_r_eq}, we can write,
\begin{align}
\langle &+z^{(q_1)}|x|+z^{(q_2)}\rangle=\sum_{\bg_1,\bg_2}C_+^*(\bg_1)C_+(\bg_2)\nonumber\\
&\times\int e^{-i(\bg_1-\bg_2).\textbf{r}}\Psi_{xyz}^{-z,(q_1)}x\Psi_{xyz}^{+z,(q_2)}d\textbf{r},
\end{align}
\begin{figure}[b!]
\begin{center}
\includegraphics[width=.48\textwidth]{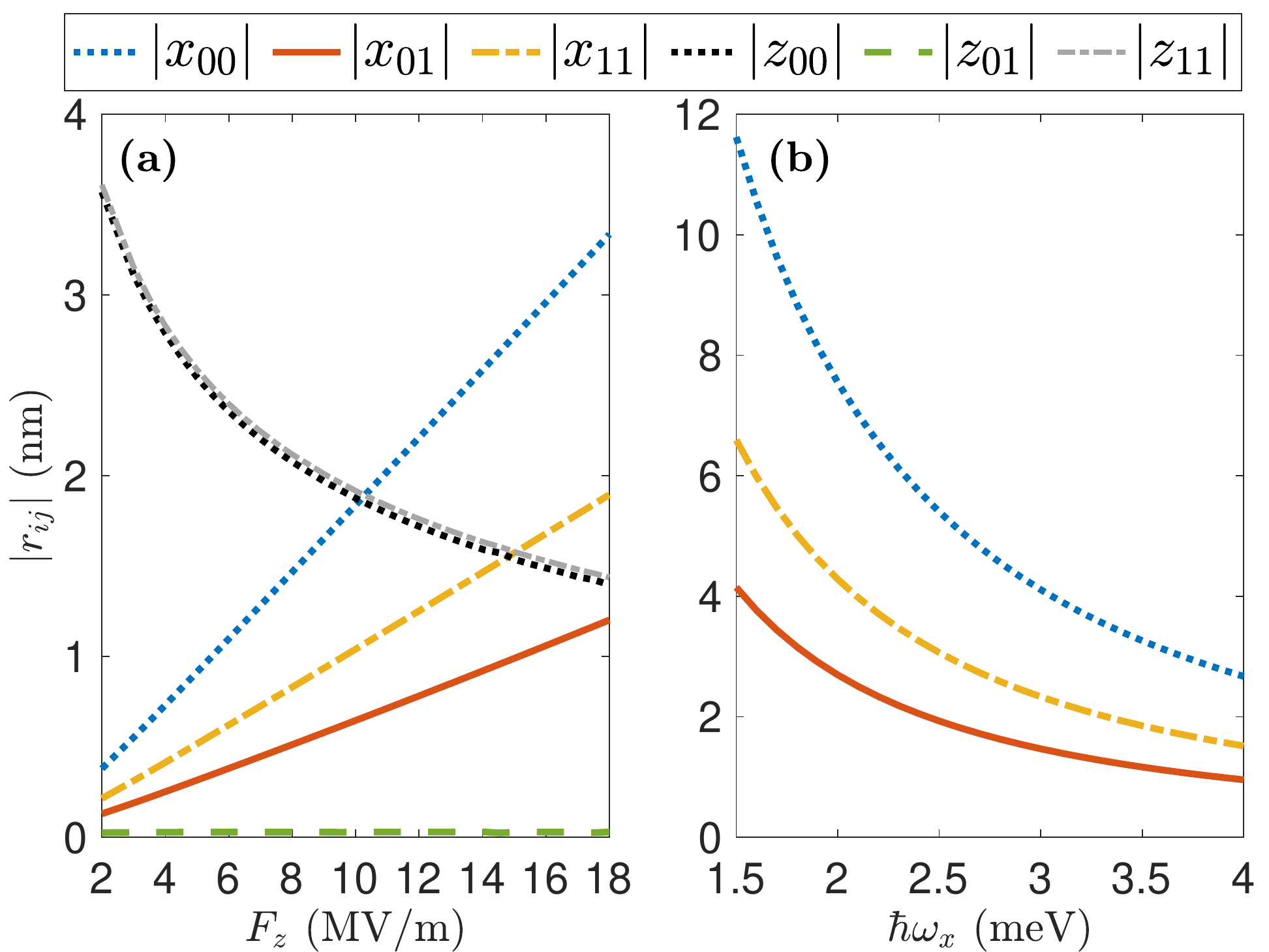}
\end{center}
\caption{(a) The dipole matrix elements as a function of the out-of-plane electric field. (b) The in-plane dipole matrix elements as a function of the in-plane orbital splitting at $B=0$. In both panels, all the other parameters are the same as used in Fig. \ref{fig:DP_PD}.  }  \label{fig:r_ij_vs_F_z_Ex}
\end{figure}

The above equation indicates that the only wavevectors that contribute to the sum are the ones where $\bg_1=\bg_2$. It is then easy to show  $\sum_{\bg}C_+^*(\bg)C_+(\bg)=1$ due to the normalization of the wavefunction. To calculate the integral, we note that for harmonic oscillators we can write 
\begin{align}
    x=\frac{1}{2}x_0(a+a^\dagger),
\end{align}
where $a$ and $a^\dagger$ are the ladder operators. Using this together with the valley-dependent envelope function \eref{eq:Psi_t_p}, we readily find
\begin{align}\label{eq:+zx+z}
    \langle &+z^{(q_1)}|\hat{x}|+z^{(q_2)}\rangle=\frac{1}{2}x_0\left(c_{1,0}^{-z,(q_1)} + c_{1,0}^{+z,(q_2)}\right).
\end{align}
In the same way, we find,
\begin{align}\label{eq:+zx+z}
    \langle &-z^{(q_1)}|\hat{x}|-z^{(q_2)}\rangle=\frac{1}{2}x_0\left(c_{1,0}^{+z,(q_1)} + c_{1,0}^{-z,(q_2)}\right).
\end{align}
If $q_1\neq q_2$, from the above two equation and \eref{eq:O_qq'}, it is easy to arrive to \eref{eq:ME_x_nd}. Similarly, when $q_1 = q_2$ and given \eref{eq:O_qq} it is easy to verify \eref{eq:ME_x_d}. The dipole matrix elements of $z$ given by \eref{eq:ME_z_nd} and \eref{eq:ME_z_d} are also derived in the same way by noting that the contribution from the Block periodic part of the wave function average to 1, as presented here.

In Fig.~\ref{fig:r_ij_vs_F_z_Ex}(a), we show the dipole matrix elements as a function the electric field. By increasing $F_z$, the out-of-plane confinement becomes stronger and therefore the $z$ dipole matrix elements becomes smaller.  On the other hand, increasing the electric field further pushes the envelope function towards the interface. Note that the coefficients $c_{1,0}^{\pm z, (q)}$ in \esref{eq:ME_x_nd} and (\ref{eq:ME_x_d}) originate from the envelope function at a narrow region around the interface set by the width of the step, see \esref{eq:PH_p} and (\ref{eq:FH_p}). Using \esref{eq:psi_0z} and (\ref{eq:psi_zAT0}), it is easy to justify the linear scaling of $x_{ij}$ in Fig.~\ref{fig:r_ij}(a) with respect to the electric field.

In Fig.~\ref{fig:r_ij_vs_F_z_Ex}(b) we  show the in-plane dipole matrix elements as a function of the in-plane orbital splitting at $B=0$. Note that $x_0\propto \omega_x^{-1/2}$ and $c_{1,0}^{\pm z, (q)}\propto \omega_x^{-1}$. As such, the $x$ dipole moments scale by the orbital splitting as $\omega_x^{-3/2}$. We also note that since the in-plane magnetic field further confines the in-plane envelope function, according to \eref{eq:lB}, $\textbf{B}_{||}$ also modifies the in-plane dipole moments. However, up to a few Tesla, the magnetic confinement length \eref{eq:lB} remains small (e.g. at $B=2$ T, we find $l_B\simeq 18.1$ nm) and the change in $|x_{ij}|$ is therefore negligible. 

\section{Spin-valley coupling}\label{app:ME_SOI}
Here we present in detail the expression we find for the inter-valley spin-valley coupling, $\Delta_{32}=\langle 3|H_R + H_D|2 \rangle$ and intra-valley coupling, $\Delta_{21}=\langle 2|H_R + H_D|1 \rangle$ discussed in Sec.~\ref{sec:SQL}. The form of the interface-induced spin-orbit interaction in the presence of interface steps is given in \esref{eq:H_R} and (\ref{eq:H_D}) for the Rashba and Dresselhaus terms. Let us first define $F_R=\mathcal{S}_\mathrm{int}(x,z)$ and $F_D=\cos(\frac{4\pi z}{a_0})\mathcal{S}_\mathrm{int}(x,z)$ for convenience. We can then write,
\begin{align}
    \Delta_{32} &= \frac{1}{2}\gamma_R\sigma_x^{\downarrow\uparrow}\langle\nu^{(q=1)}|\{p_y,F_R\}|\nu^{(q=0)}\rangle\nonumber\\
    &-\frac{1}{2}\gamma_R\sigma_y^{\downarrow\uparrow}\langle\nu^{(q=1)}|\{p_x,F_R\}|\nu^{(q=0)}\rangle\nonumber\\
    &+\frac{1}{2}\gamma_D\sigma_x^{\downarrow\uparrow}\langle\nu^{(q=1)}|\{p_x,F_D\}|\nu^{(q=0)}\rangle\nonumber\\
    &-\frac{1}{2}\gamma_D\sigma_y^{\downarrow\uparrow}\langle\nu^{(q=1)}|\{p_y,F_D\}|\nu^{(q=0)}\rangle.
\end{align}
To calculate the matrix elements shown above, we use \esref{eq:O_qq'} and (\ref{eq:pmZ}) and find for both $q=q'$ and $q\neq q'$,
\begin{align}\label{eq:ME_py}
        \langle\pm (\mp)z^{(q)}|\{p_y,F_{R}\}|\pm z^{(q')}\rangle=-2B_xf_\alpha,
\end{align}
in which $f_\alpha$ is defined as
\begin{align}\label{eq:falpha}
    f_\alpha&=
    \frac{2\hbar}{y_0'}\sum_{\bg_1,\bg_2}\mathcal{G}_{++}(\bg_1,\bg_2)\nonumber\\
    &\times\sum_{n=1}^{n=3}\alpha_n\int\psi_{x,0}^2\psi_{z,0}\psi_{z,n}F_{R}e^{i(G_{2z}-G_{1z})z}dxdz,
\end{align}
where,
\begin{align}\label{eq:G++}
    &\mathcal{G}_{++}(\bg_1,\bg_2)=\nonumber\\&C^{*}_{+}(\bg_1)C_{+}(\bg_2)\delta(G_{2x}-G_{1x})\delta(G_{2y}-G_{1y}).
\end{align}
As such, using \esref{eq:O_qq'} and (\ref{eq:ME_py}) we arrive to 
\begin{align}
    \frac{1}{2}\langle\nu^{(q=1)}|\{p_y,F_R\}|\nu^{(q=0)}\rangle = -iB_xf_\alpha\sin\phi_v
\end{align}
The matrix element of $\{p_y,F_D\}$ is also found in the similar way, where in this case we find $f_\alpha\rightarrow f_\alpha'$ in which $f_\alpha'$ is defined similar to \eref{eq:falpha} by replacing $F_R\rightarrow F_D$. It is easy to prove that $f_\alpha$ and $f_\alpha'$ are real quantities.

We also find,
\begin{align}\label{eq:me_py_v1v0}
     \langle&+z^{(q=1)}|\{p_x,F_{R}\}|+z^{(q=0)}\rangle=2B_yf_\beta + 2g_1,
\end{align}
where here we defined,
\begin{widetext}
\begin{align}\label{eq:fbeta}
        f_\beta&=\frac{\hbar }{x_0'}\sum_{\bg_1,\bg_2}\mathcal{G}_{++}(\bg_1,\bg_2)\sum_{n=1}^{n=3}\beta_n\int\left(\psi_{x,0}^2+\psi_{x,1}^2-\sqrt{2}\psi_{x,0}\psi_{x,2}\right)\psi_{z,0}\psi_{z,n}F_{R}e^{i(G_{2z}-G_{1z})z}dxdz,\\
        g_1&=\frac{i}{2}\frac{\hbar}{x_0'}\sum_{\bg_1,\bg_2}\mathcal{G}_{++}(\bg_1,\bg_2)\sum_{(m,n)\neq(0,0)}\left[c_{m,n}^{+z,(0)}-c_{m,n}^{-z,(1)}\right]\nonumber\\
    &\times\int\left(\sqrt{m+1}\psi_{x,m+1}\psi_{x,0}-\psi_{x,1}\psi_{x,m}-\sqrt{m}\psi_{x,m-1}\psi_{x,0}\right)\psi_{z,0}\psi_{z,n}F_{R}e^{i(G_{2z}-G_{1z})z}dxdz,
\end{align}
\end{widetext}
where we used the time-reversal symmetry relation $c_{m,n}^{ -z,(q)}=[c_{m,n}^{+z,(q)}]^*$. We further find (noting that $f_\beta$ is a real quantity), 
\begin{align}
     \langle-z^{(q=1)}|\{p_x,F_{R}\}|-z^{(q=0)}\rangle&=2B_yf_\beta - 2g_1^*,\\
     \langle+z^{(q=1)}|\{p_x,F_{R}\}|-z^{(q=0)}\rangle&=2B_yf_\beta+ 2g_2,\\
     \langle-z^{(q=1)}|\{p_x,F_{R}\}|+z^{(q=0)}\rangle&=2B_yf_\beta -2g_2^*,
\end{align}
in which $g_2$ reads
\begin{widetext}
\begin{align}\label{eq:D_px}
g_2= \frac{i}{2}\frac{\hbar}{x_0'}&\sum_{\bg_1,\bg_2}C_{+}(G_1)C_{-}(G_2)\delta(G_{1x}-G_{2x})\delta(G_{1y}-G_{2y})\sum_{(m,n)\neq(0,0)}\left[c_{m,n}^{-z,(0)}-c_{m,n}^{-z,(1)}\right]\nonumber\\
    &\quad\quad\times\int\left(\sqrt{m+1}\psi_{x,m+1}\psi_{x,0}-\psi_{x,1}\psi_{x,m}-\sqrt{m}\psi_{x,m-1}\psi_{x,0}\right)\psi_{z,0}\psi_{z,n}F_{R}e^{i(G_{2z}-G_{1z}-2k_0)z}dxdz\Bigg].
\end{align}
\end{widetext}
Using the above equations, we arrive to 
\begin{align}\label{eq:me_px_v1v0}
    \frac{1}{2}\langle\nu^{(q=1)}|\{p_x,F_R\}|\nu^{(q=0)}\rangle = g_c+iB_yf_\beta\sin\phi_v
\end{align}
in which we defined,
\begin{align}\label{eq:gc}
    g_c = \Re[g_1 - e^{-i\phi_v}g_2].
\end{align}
The matrix element of $\{p_x,F_D\}$ is calculated in the similar way by replacing $f_\beta\rightarrow f_\beta'$ and $g_c\rightarrow g_c'$ in which these two quantities are defines by replacing $F_R\rightarrow F_D$ in all related relations presented above. 
We note that the Pauli matrices in the spin-orbit interaction are defined with respect to the lattice crystallographic axes whereas the spin states are defined with respect to the direction of the applied magnetic field. For an in-plane magnetic field, we then have $\sigma_x^{\uparrow\uparrow}=-\sigma_x^{\downarrow\downarrow}=\cos\phi_B$, $\sigma_x^{\downarrow\uparrow}=-i\sin\phi_B$,  $\sigma_y^{\uparrow\uparrow}=-\sigma_y^{\downarrow\downarrow}=\sin\phi_B$ and $\sigma_y^{\downarrow\uparrow}=i\cos\phi_B$. Using these relations together with \esref{eq:me_py_v1v0} and (\ref{eq:me_px_v1v0}), it is easy to arrive the expression for $\Delta_{32}$ given by \eref{eq:D_32}. 

 On the other hand, we find for the intra-valley matrix elements,
 \begin{align}\label{eq:me_vqvq}
    \frac{1}{2}\langle\nu^{(q)}|\{p_x,F_R\}|\nu^{(q)}\rangle &= B_yf_\beta\left(1+\cos\phi_v\right),\\
    \frac{1}{2}\langle\nu^{(q)}|\{p_y,F_R\}|\nu^{(q)}\rangle &= -B_xf_\alpha\left(1+\cos\phi_v\right).
\end{align}
It is then easy to verify the form of the intra-valley spin-orbit coupling $\Delta_{21}$ given by \eref{eq:D_21}.

\section{Corrections due to SOM}\label{app:som}
Here we present the relaxation for the coefficients $c_1$ to $c_4$ in \esref{eq:ql_SOM_g} and (\ref{eq:ql_SOM_e}) that give the corrections to the spin qubit levels due to spin-orbit mixing. Given the interface-induced spin-orbit interaction \esref{eq:H_R_I} and (\ref{eq:H_D_I}), we find by using the standard perturbation theory at the first order,
\begin{align}
    c_1&=\frac{1}{2}\gamma_R\frac{\langle 1_x|\{p_x,F_R\}|0\rangle}{\hbar\omega_x'+E_z}\sigma_y^{\uparrow\downarrow}-\frac{1}{2}\gamma_D\frac{\langle 1_x|\{p_x,F_D\}|0\rangle}{\hbar\omega_x'+E_z}\sigma_x^{\uparrow\downarrow},\\
     c_2&=\frac{1}{2}\gamma_D\frac{\langle 1_y|\{p_y,F_D\}|0\rangle}{\hbar\omega_y'+E_z}\sigma_y^{\uparrow\downarrow}-\frac{1}{2}\gamma_R\frac{\langle 1_y|\{p_y,F_R\}|0\rangle}{\hbar\omega_y'+E_z}\sigma_x^{\uparrow\downarrow}.
\end{align}
The coefficient $c_3$ ($c_4$) is defined similar to $c_1$ ($c_2$) by replacing $E_z\rightarrow-E_z$ and $\sigma_{x(y)}^{\uparrow\downarrow}\rightarrow\sigma_{x(y)}^{\downarrow\uparrow}$.  Here $F_{R}$ and $F_{D}$ are the same as defined in Appendix~\ref{app:ME_SOI}, and the states $|0\rangle$, $|1_x\rangle$ and $|1_x\rangle$ are defined in Sec.~\ref{sec:SQL_SOM}. We also find,
\begin{widetext}
\begin{align}\label{eq:px_1x0}
    \langle 1_x|\{p_x,F_{R(D)}\}|0\rangle&=\frac{i\hbar}{x_0'}\sum_{\bg_1\bg_2}\mathcal{G}_{++}\int (\psi_{x,0}^2+\psi_{x,1}^2-\sqrt{2}\psi_{x,0}\psi_{x,2})\psi_{z,0}^2F_{R(D)}e^{-i(G_{1z}-G_{2z})z}dxdz,\\\label{eq:py_1y0}
    \langle 1_y|\{p_y,F_{R(D)}\}|0\rangle&=\frac{2i\hbar}{y_0'}\sum_{\bg_1\bg_2}\mathcal{G}_{++}\int \psi_{x,0}^2\psi_{z,0}^2F_{R(D)}e^{-i(G_{1z}-G_{2z})z}dxdz,
\end{align}
\end{widetext}
where $\mathcal{G}_{++}$ is defined in \eref{eq:G++}.

\section{Dipole moments between the modified qubit states}

\begin{figure}[t!]
\begin{center}
\includegraphics[width=.48\textwidth]{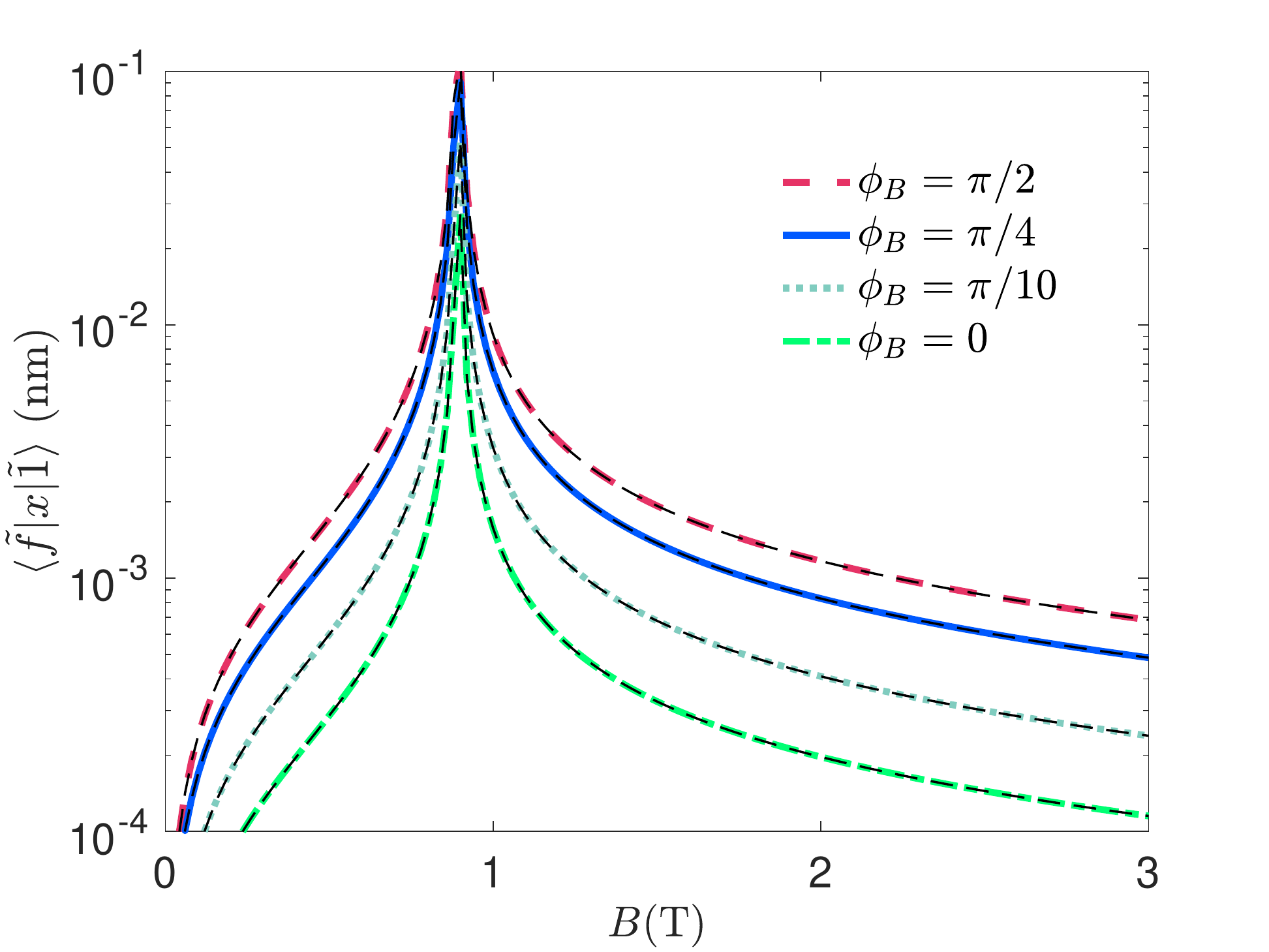}
\end{center}
\caption{The dipole matrix elements between modified qubit levels due to SVM as a function of the magnetic field for various $B$ directions, $\phi_B$. The color lines are obtained by only considering the inter-valley coupling $\Delta_{32}$ in finding the qubit modified levels, see, \eref{eq:r_q_21} for the dipole moment below the hotspot. The black dashed lines are obtained by considering both inter-valley and intra-valley couplings, $\Delta_{32}$ and $\Delta_{21}$. All other parameters are the same as used in Fig.~\ref{fig:T_1_the_vs_exp}. }  \label{fig:r_q_ij}
\end{figure}

Here we present the dipole matrix elements between the modified qubit levels. We begin by considering the SVM in which case, by neglecting the intra-valley coupling $\Delta_{21}$, the qubit levels are given by \esref{eq:qulv_1}, (\ref{eq:qulv_2}) and (\ref{eq:qulv_3}). We find for $r=(x,z)$,
\begin{align}\label{eq:r_q_21}
    \langle\tilde{2}|r|\tilde{1}\rangle &= \frac{1}{2}\langle \nu^{(q=1)}|r|\nu^{(q=0)}\rangle \\
    &\times\left[-\sqrt{(1+a_{-})(1+a_{+})}+\sqrt{(1-a_{-})(1-a_{+})}\right],\nonumber
\end{align}
where we used $\langle \nu^{(q=0)}|r|\nu^{(q=1)}\rangle=-\langle \nu^{(q=1)}|r|\nu^{(q=0)}\rangle$ as a result of \esref{eq:ME_x_nd} and (\ref{eq:ME_z_nd}). The $\langle\tilde{3}|r|\tilde{1}\rangle$ dipole moment can also be found in a similar way. In Fig.~\ref{fig:r_q_ij} we show the $\hat{x}$ dipole moment between modified qubit levels. As noted in Sec.~\ref{sec:SQL_SVM}, below the hotspot, the logical qubit excited state is $|\tilde{f}\rangle=|\tilde{2}\rangle$ whereas above the hotspot the logical qubit excited state is $|\tilde{f}\rangle=|\tilde{3}\rangle$. 
The black dashed line in the figure are obtained by considering the intra-valley coupling $\Delta_{21}$ as well and finding the qubit levels by exact diagonalization. 

We observe that the effect of the intra-valley coupling is indeed negligible as mentioned in Sec.~\ref{sec:SQL_SVM}. The anisotropic behavior of the dipole moment is due to the anisotropic response of the inter-valley spin-valley coupling $\Delta_{32}$, see Fig.~\ref{fig:D_32}. 

\label{app:dip_qlel}

\end{document}